\documentclass[onecolumn,showpacs,amsfonts,aps,prc,nofootinbib,floatfix,%
superscriptaddress]{revtex4}

\usepackage{amsmath}
\usepackage{bm}
\usepackage{graphicx}

\newcommand{\tg}{\tan}

\usepackage{epsfig}
\newcommand{\beq}{\begin{equation}}
\newcommand{\eeq}{\end{equation}}
\newcommand{\bea}{\vspace{0.25cm}\begin{eqnarray}}
\newcommand{\eea}{\end{eqnarray}}

\newcommand{\ta}{\mbox{{\boldmath
$\tau$}}}
\newcommand{\delb}{\mbox{{\boldmath
$\delta$}}}

\newcommand{\tb}{\mbox{{\boldmath
$\tau$}}}

\newcommand{\ro}{\mbox{{\boldmath
$\rho$}}}

\newcommand{\kb}{\mbox{{\bf
k}}}
\newcommand{\qb}{\mbox{{\bf
q}}}
\newcommand{\pb}{{{\bf p}}}
\newcommand{\rb}{\mbox{{\bf
r}}}

\newcommand{\bb}{{{\bf b}}}
\newcommand{\eb}{{{\bf e}}}

\newcommand{\ub}{{{\bf u}}}

\newcommand{\Ab}{{{\bf A}}}

\newcommand{\Rb}{{{\bf R}}}

\newcommand{\pt}{\mbox{{\bf
p}}_\perp}

\def\lsim{\mathrel{\rlap{\lower4pt\hbox{\hskip1pt$\sim$}}
    \raise1pt\hbox{$<$}}}         
\def\gsim{\mathrel{\rlap{\lower4pt\hbox{\hskip1pt$\sim$}}
    \raise1pt\hbox{$>$}}}         

\newcommand{\landau}{L.D.~Landau Institute for Theoretical Physics,
        GSP-1, 117940, Kosygina Str. 2, 117334 Moscow, Russia}

\begin{document}


\title{Radiative contribution to 
$p_{\perp}$-broadening of fast partons in a quark-gluon plasma
}
\date{\today}

\author{B.G.~Zakharov}\affiliation{\landau}

\begin{abstract}
The  contribution  of  radiative  processes  to   
$p_{\perp}$-broadening  of  fast  partons  in  a  quark-gluon
plasma is investigated. Calculations are performed beyond the soft gluon 
approximation. It is shown that the radiative correction to 
$\langle p_{\perp}^2\rangle$ for conditions of heavy ion collisions at 
RHIC and LHC is negative and can be
comparable in absolute value with the nonradiative contribution. 
This prediction differs radically from the
essentially positive contribution of radiative processes to 
$p_{\perp}$-broadening, which was predicted earlier in the literature.
\end{abstract}
%

\maketitle

\section{Introduction}
The results of experiments on collision of relativistic heavy nuclei at
RHIC and LHC provide a 
great deal of
evidence  on  the  formation  in  the  initial  stage  of
nuclear collisions of a hot QCD matter in the quark-gluon  plasma  (QGP)  phase.  The  QGP  formation  is
confirmed by successful simulation of 
$AA$ collisions in
hydrodynamic models that require the formation at  the proper
time  $\tau\sim 0.5-1$ fm \cite{Heinz_hydro,Heinz_tau,Pasechnik} of a
medium with a temperature $2-4$ times higher than the
deconfinement  temperature  $T_c\approx 160$ MeV.
The suppression of the spectra   of   particles   with   large   
transverse   momenta
observed  in  experiments  on  
$AA$ collisions,  which  is
characterized  by  the nuclear  modification  factor  
$R_{AA}$, is
also considered as a signal of the QGP formation 
\cite{Wied_JQ,Pasechnik}.
It is generally accepted that the suppression of particle spectra, which is quite strong for RHIC and LHC
($R_{AA}\sim 0.1-0.2$ in  central  collisions  for  particles  with
$p_{\perp}\sim 10-20$ GeV) is associated with the jet modification  due  to  
collisional  \cite{Bjorken1}  and  radiative  energy  losses
\cite{GW,BDMPS1,BDMPS2,LCPI1,GLV1,AMY,W1} of fast partons in the QGP. This modification
of jets in the QGP is usually referred to as jet quenching (JQ) in the literature. For RHIC and LHC conditions, the dominant contribution to energy loss comes
from the radiative mechanism of induced gluon emission 
\cite{Z_coll,Gale_coll}. Induced gluon emission is caused by parton multiple 
scattering in the medium. For the RHIC
and LHC conditions, induced gluon emission off fast
quarks and gluons is an essentially collective process,
in which, like that in photon emission by electrons in
a  conventional  matter,  multiple  scatterings,  leading
to  the  Landau-Pomeranchuk-Migdal  suppression,
play   an   important   role \cite{LP,Migdal}.   The   available
approaches  to  radiative  energy  loss  and  to  the  
Landau-Pomeranchuk-Migdal effect in QCD are based 
on  the  approximation of  one-gluon emission \cite{GW,BDMPS1,BDMPS2,LCPI1,GLV1,AMY,W1}.  
The  induced spectrum of gluon emission by a fast parton in a
medium  can  be  expressed  via  the  solution  to  the  2D
Schr\"odinger equation with an imaginary potential \cite{LCPI1,BDMPS1},
which  can  be  expressed  via  the  product  of  the  QGP
number density and the dipole cross section $\sigma_{q\bar{q}}(\rho)$ of 
scattering of
a $q\bar{q}$ pair off the QGP constituent (here, 
$\rho$ is the size of
a  $q\bar{q}$  pair).  In  the  quadratic  approximation  $\sigma_{q\bar{q}}(\rho)\approx C\rho^2$, the induced gluon spectrum can be expressed in
terms  of  the  Green  function  of  a  harmonic  oscillator
with  a  complex  frequency.  In  the  oscillator  approximation, the square of the frequency is proportional to
the well-known transport coefficient $\hat{q}$
\cite{BDMPS1,BDMPS2} defined by the
relation   $\hat{q}=2Cn$,  where  
$n$ is  the  number density  of  the medium.

In  analysis  of  the  JQ  phenomenon,  multigluon  processes
must  also  be  considered.  However, 
even in the  simplified  oscillator  approximation \cite{Z_OA},
 the  inclusion  of
multigluon processes is a complicated problem 
\cite{Arnold_2g1}.
At present, the emission of several gluons is usually taken
into account in the approximation of the independent
gluon emission  \cite{BDMS_RAA}.   In  this  approximation,  it  is
possible to reach reasonable agreement with the RHIC
and  LHC  data  on  nuclear  modification  factors  
$R_{AA}$  \cite{RAA13,RPP14}. 
Since energy losses for partons substantially
depend on the number density of the medium, analysis of the
data  on  
$R_{AA}$ is  an  effective  tool  for  diagnostics  of  the
QGP formed in 
$AA$ collisions. In calculations of radiative  energy  loss  in  the
oscillator  approximation,  the
data on 
$R_{AA}$  provide information on the value of $\hat{q}$   in a
plasma fireball and, hence, on the QGP density. It is
important  that  despite  the  approximate  nature  of
modern approaches to JQ, the entropy/energy density
of the QGP required for concordance with the RHIC
and LHC data on 
$R_{AA}$ is in reasonable agreement with
the results obtained in the hydrodynamic models of 
$AA$ collisions.

Apart from modification of the jet longitudinal structure,  
which  leads  to  suppression  of  particle
spectra, rescatterings of fast partons in the QGP must
also change the direction of the jet. For an individual
parton, the intensity of variation of its transverse (relative to the
direction of the velocity of 
the initial parton)  momentum  $p_\perp$
due  to  multiple  scattering  in  the
medium in the oscillator approximation is characterized by the same transport
coefficient $\hat{q}$ \cite{BDMPS2}, which also
determines the induced gluon emission. For a parton
traversing a homogeneous medium, the mean square
of the transverse momentum is given by
\beq
\langle p_{\perp}^2\rangle=\hat{q}L\,,
\label{eq:10-1}
\eeq 
where $L$ is the path length in the medium. The Coulomb effects that are lost
in the quadratic approximation  lead  to  
a  slight  (logarithmic)  deviation  from
purely  linear  dependence  of  
$\langle p_{\perp}^2\rangle$ on $L$. Experimentally, 
$p_\perp$-broadening  of  fast  partons  can  be  manifested in an increase of azimuthal jet decorrelation in
the di-jet events (or in decorrelation of a photon and
the  jet  in  the  photon-jet  events)  in  
$AA$ collisions  as
compared  to  
$pp$  collisions.  The  observation  of  effects associated  with  
$p_\perp$-broadening  can  provide  direct
information on  the QGP fireball in 
$AA$ collisions.
For  understanding  the  JQ  mechanisms,  it  would  be
interesting to compare the values of $\hat{q}$   extracted from
the 
$R_{AA}$ data with that obtained from the results on the jet 
$p_\perp$-broadening.
The  experimental  detection  of  the jet $p_\perp$-broadening  is complicated by the fact that strong azimuthal jet
decorrelation effects occur even for 
$pp$ collisions due to
the  Sudakov  form  factors \cite{Mueller_dijet}.  For  this  reason,  the
observation of the jet $p_\perp$-broadening induced by interaction  with  
the  QGP  requires  measurements  with  a
high degree of precision. The available data at RHIC \cite{STAR1}
and LHC energies \cite{ALICE_hjet} do not allow to draw a
definite conclusion on the jet $p_\perp$-broadening  in the
QGP. Nevertheless, it is expected that after improving
the accuracy of the data, it will be possible to observe
the jet $p_\perp$-broadening \cite{Gyulassy_dijet}.

One of the important theoretical problems arising
in connection with $p_\perp$-broadening of jets in the QGP (as
well  as  with  the  JQ  phenomenon)  is  the  problem  of
contribution to $p_\perp$-broadening of  radiative  corrections  due  to  the  soft
gluon  emission  \cite{Wu,Mueller_pt,Blaizot_pt}.  It  was
expected that the recoil effects in the emission of soft
gluons must enhance $p_\perp$-broadening. Since the formation length of soft gluons is small, this effect can be
treated as local in the longitudinal coordinate and can
be  interpreted  as  renormalization  of $\hat{q}$.  In \cite{Mueller_pt} it  was
found that the main contribution to the radiative correction  to  
$\langle p_{\perp}^2\rangle$ 
for  a  homogeneous  QGP  has  a double
logarithmic form
\beq
\langle p_{\perp}^2\rangle_{rad}\sim \frac{\alpha_sN_c\hat{q}L}{\pi}\ln^2(L/l_0)\,,
\label{eq:20-1}
\eeq
 where $l_0$  is the size on the order of the Debye radius in
the QGP. For typical parton path length $L\sim 5$ fm in
the  QGP  for  central  collisions  of  heavy  nuclei,  the
radiative contribution to $\langle p_{\perp}^2\rangle$
 turns out to be comparable  with  conventional  nonradiative  
contribution (\ref{eq:10-1}).
In \cite{Mueller_pt}, a generalization of the approach of \cite{LCPI1} for the
induced gluon energy spectrum to the case of the double
differential  spectrum  on  transverse  momentum  and
energy has been used for calculating the radiative contribution to 
$p_\perp$-broadening. It should be noted that
the  corresponding  expressions  have been  obtained  without
using the soft gluon approximation in our earlier work \cite{LCPI_PT}
(see   also  \cite{BSZ,Z_NP05}),   which   was   apparently
unknown  to  the  authors  of \cite{Mueller_pt}.  In  the  soft  gluon
approximation,  the  induced  gluon  spectrum  on  the
energy  and  the  transverse  momentum  have  also  been
considered in \cite{W1}.

In this study, using the technique of \cite{LCPI1}, in the form
developed  in \cite{LCPI_PT}  for  the  spectrum  in  the  Feynman
variable  and  transverse  momentum  for  the  induced
$a\to bc$
 transition in the medium, we address the radiative contribution to
 $p_\perp$-broadening beyond the soft
gluon approximation (the formalism developed in \cite{LCPI1,LCPI_PT} will 
be referred to as the light-cone path integral
(LCPI) approach). It will be shown that, in this case,
there  appear  no  double  logarithmic  terms  associated
with  rescatterings  of  the  initial  parton  in  the  QGP,
which make a negative contribution to $p_\perp$-broadening so that the total contribution  turns out to be
negative for the RHIC and LHC conditions. In contrast  to  the  double
logarithmic  contribution  
considered in \cite{Mueller_pt}, this contribution is not local and cannot
be interpreted as a renormalization of the transport coefficient $\hat{q}$. As in
\cite{Mueller_pt}, we analyze a homogeneous QGP in
the oscillator approximation.

The paper is organized as follows. In Section 2, we
review  (for  convenience  of  the  reader)  the  LCPI
method  for  calculating  the  double  differential  spectrum  in  the longitudinal  Feynman  variable  
$x$ and  in  the
transverse momentum for the induced 
$a\to bc$ transitions.  In  Section  3,  the  calculation  of  the  radiative
contribution to $p_\perp$-broadening is discussed. In Section 4, numerical
results for the RHIC and 
LHC conditions  are  considered.  Conclusions  are  contained  in Section 5. 
Some expressions referring to our calculations are given in two appendices.

\section{SPECTRUM  OF  THE  INDUCED  
$a\to bc$ TRANSITION  IN  THE  LCPI  METHOD}
For  the  convenience  of  the  reader,  we  briefly
describe in this section the basic concepts of the LCPI
formalism  \cite{LCPI1,LCPI_PT} for  processes  of  type $a\to bc$ 
  in  an
amorphous   medium.   In   the   LCPI   approach,   we
assume  that  the  energies  of  all  particles  are large  as
compared  to  their  masses.  We  also  assume  that  the
transverse momenta of particles are small as compared
to their energies; i.e., we perform analysis in the small-angle approximation 
(angles are determined relative to
the direction of the momentum of the initial particle
$a$). This approximation is very good for radiative processes  
at  high  energies  in  QED  \cite{LL4}.  It  remains  good
enough for processes with fast partons in QCD matter
also  \cite{Z_kinb}.   In   the   LCPI   approach,   the   difference
between final expressions for the transition probability
of the $a\to bc$
 process in the Abelian and non-Abelian
cases is found to be minimal. For relativistic particles,
spin effects in the interaction of particles with matter
can be disregarded, and multiple rescatterings of particles 
in the matter occur in the same way as for scalar
particles. Spin effects are manifested only in the emergence of 
vertex operators for the $a\to bc$
 transition and
have the form analogous to that for such transitions in
vacuum.   The   evolution   of   wavefunctions   in   the
medium before and after splitting 
$a\to bc$ in the leading
order  in  the  particle  energy  approximation,  is  
independent  of  spin  factors.  Therefore, for  simplicity,  we  will  
illustrate  
the  formalism  for  the $a\to bc$  transition  in  the
electromagnetic field of an amorphous medium in the
case of spinless particles with the Lagrangian of interaction between 
fields of 
$a$, $b$, and $c$
\beq
L_{int}=\lambda \hat{\psi}_b^{+}\hat{\psi}_b^{+}\hat{\psi}_a+\mbox{(h.c.)}\,. 
\label{eq:30-2}
\eeq

\subsection{The $a\to bc$ transition in a medium
for scalar particles}
We assume that the 
$z$
 axis is chosen in the direction
of the momentum of initial particle 
$a$
 prior to its interaction with the medium; the matter occupies a 
finite region $0 < z < L$,
 and is homogeneous in the transverse
coordinates.  The  element  of  the   $\hat{S}$-matrix  for  the
induced $a\to bc$
  transition  for  the Lagrangian  (\ref{eq:30-2})  in  the
field of the medium can be written in the form
\beq
\langle bc|\hat{S}|a\rangle=i\int\! dt
d\rb \lambda
\psi_{b}^{*}(t,\rb)\psi_{c}^{*}(t,\rb)\psi_{a}(t,\rb)\,,
\label{eq:40-2}
\eeq
where $\psi_{i}$  are  the  wavefunctions  of  particles  in  the
external field of the medium. Each of the initial wavefunctions $\psi_{i}$
 satisfies the Klein-Gordon equation
\beq
[(\partial_{\mu}+ie_{i}A_{\mu})(\partial^{\mu}+ie_{i}A^{\mu})+m_{i}^{2}]
\psi_{i}(t,\rb)=0\,,
\label{eq:50-2}
\eeq
where $e_i$
 is the particle charge. Let us first consider the
initial particle impinging on the medium from infinity.
In this case, for particle $a$, we must choose an appropriate 
wavefunction which has the plane wave form at
$z\to -\infty$,  while,  for  the  final  
$b$  and  $c$ particles,  we
choose  outgoing  wavefunctions  in  the  form  of  plane
waves for $z\to \infty$. We assume that $m_a<m_b+m_c$; therefore,  
there  is  no  transition  $a\to bc$   in  vacuum.  The
wavefunctions of fast particles for $E_{i}\gg m_{i}$ are rapidly
oscillating functions of variables $t$ and $z$.
Therefore, it is convenient to write $\psi_{i}$ in the form
\beq
\psi_{i}(t,\rb)=\frac{1}{\sqrt{2E_{i}}}\exp[-iE_{i}(t-z)]
\phi_{i}(t,\rb)\,,
\label{eq:60-2}
\eeq
where $\rb=(z,\ro)$, $\ro$ being the transverse coordinate. As
usual, we normalize the fluxes for free plane waves to
unity, which corresponds to 
$|\phi_{i}|=1$ at $z\to -\infty$ for $i=a$ 
and  at $z\to\infty$ for $i=b,c$. Obviously, in expression (\ref{eq:60-2}),
the dependence of $\phi_{i}$ (these functions will be referred
to as transverse wavefunctions) on $t$
 and on the longitudinal coordinate $z$ must be smooth. For the case of
time-independent external potential, transverse wavefunctions 
$\phi_{i}$ are independent of $t$
 and are functions of the  longitudinal  coordinate  $z$
  and  of  the  transverse vector $\ro$. In  this  case,  after  
integrating  over  $t$,  we  can
single  out  in  the  $\hat{S}$-matrix  element  the $\delta$-function  of
the energy difference and write it in terms of the integral over 
the spatial variables
\beq
\langle bc|\hat{S}|a\rangle=
\frac{i2\pi\delta(E_{b}+E_{c}-E_{a})}{\sqrt{8E_aE_bE_c}}\int_{z_i}^{z_f}\! dz\int
d\ro \lambda
\phi_{b}^{*}(z,\ro)\phi_{c}^{*}(z,\ro)\phi_{a}(z,\ro)\,,
\label{eq:70-2}
\eeq
where $z_i=-\infty$ and $z_f=\infty$.

Using the Fermi golden rule, one can obtain from
relation (\ref{eq:70-2}) the  following  expression  for  the  differential 
probability of transition $a\to bc$, averaged over the
states of the target
\beq
\frac{dP}{dx d\qb_{b}d\qb_{c}}=\frac{2}{(2\pi)^{4}}
\mbox{Re}\!
\int\!d\ro_{1}d\ro_{2}
\int_{z_{1}<z_{2}}dz_{1}dz_{2}\,
\hat{g}\langle 
W(z_{1},\ro_{1})
W^{*}(z_{2},\ro_{2})
\rangle\,,
\label{eq:80-2}
\eeq
where
$W(z,\ro)=\phi_{b}^{*}(z,\ro)
\phi_{c}^{*}(z,\ro)\phi_{a}(z,\ro)$,
$\qb_{b,c}$  are  the  transverse  momenta  of  particles  
$b$  and  $c$
(note  that  we  will  use  bold  letters  only  for  transverse
vectors), $x=x_{b}=E_b/E_a$
 is the Feynman variable for
particle $b$ (since  $E_b+E_c=E_a$,  we  can  also  use
$x=x_c=E_c/E_a$  as  the  longitudinal  variable).  Symbol
$\langle ...\rangle $ in relation (\ref{eq:80-2}) indicates averaging over 
the states of the target and $\hat{g}$ indicates the vertex factor
\beq
\hat{g}=\frac{\lambda^{2}}{16\pi x_b x_c E_{a}^{2}}\,.
\label{eq:90-2}
\eeq
As will be shown below, in real QED and QCD, this
factor is a differential operator. It should be noted that
in contrast to element of the $\hat{S}$-matrix (\ref{eq:70-2}), in 
the integral with respect to $z_{1,2}$
 in formula (\ref{eq:80-2}), regions 
$|z_{1,2}|\to \infty$ can be significant. For evaluating the contribution
from  these  regions  correctly  in  the  calculation  of  the
probability  of  transition $a\to bc$,  it  is  convenient  to
assume  that  interaction  (\ref{eq:30-2})  in  Eq. (\ref{eq:90-2})  is  
switched  off adiabatically at $z\to \pm \infty$.
In this case, in Eq. (\ref{eq:90-2}), 
$\lambda^2\to \lambda(z_1)\lambda(z_2)$, where $\lambda(z)\to 0$ for 
$|z|\to \infty$.

We have not used yet the explicit form of the transverse wavefunctions. 
For $E_{i}\gg m_{i}$, after the substitution
of relation (\ref{eq:60-2}) into (\ref{eq:50-2}), we can obtain 
from relation (\ref{eq:50-2})
in leading order in energy the following equation that
describes  the  evolution  of  wavefunction  
$\phi_i(z,\ro)$ in variable $z$
\beq
i\frac{\partial{\phi_{i}}}{\partial{z}}=
\hat{H}_{i}\phi_{i}\,,
\label{eq:100-2}
\eeq
\beq
\hat{H}_{i}=\frac{(\pb_{\perp}-e_{i}\Ab_{\perp})^{2}+m_{i}^{2}}{2\mu_{i}}
+e_{i}(A^{0}-A^{3})\,,   
\label{eq:110-2}
\eeq
where $\mu_{i}=E_{i}$. For the initial particle $a$ the wavefunction $\phi_a$
 can be written as
\beq
\phi_{a}(z,\ro)=\int d\ro' K_{a}(\ro,z|\ro',z_{i})\phi_{a}(z_{i},\ro')\,.
\label{eq:120-2}
\eeq
Here $z_{i}\to -\infty$, and $\phi_{a}(z_i,\ro)\propto \exp(i\qb_{a}\ro)$ 
(the common  phase  of  the  wavefunction  is  immaterial  here),
and $K_{a}$
 is the retarded Green function for Schr\"odinger
equation (\ref{eq:100-2}) with $i=a$. 
The wavefunctions for final
particles can be expressed in terms of their values for
$z_{f}\to \infty$
and the advanced Green functions of Eq. 
(\ref{eq:100-2}) for $i=b,c$.
Using  the  fact  that  the  advanced  Green
function  is  connected  with  the  retarded  Green  function by 
the relation
$$K_{ret}(\ro_{2},t_{2}|\ro_{1},t_{1})=
K_{adv}^{*}(\ro_{1},t_{1}|\ro_{2},t_{2})\,,
$$
we can write $\phi_{b,c}(z,\ro)$ in the form
\beq
\phi_{b,c}(z,\ro)=\int d\ro' K_{b,c}^{*}(\ro',z_{f}|\ro,z)
\phi_{b,c}(z_{f},\ro')\,.
\label{eq:130-2}
\eeq
Then,  after  substituting  of  relation  
(\ref{eq:120-2}) and (\ref{eq:130-2}) into 
(\ref{eq:80-2}), the  differential  spectrum  can  be  expressed  in
terms  of  the  transverse  density  matrices  of  the  initial
particle at 
$z=z_i$ and of the final particles at $z=z_f$ 
 and the
retarded Green functions as shown in Fig. 1a. In this
diagram, the Green functions 
$K$  and the complex-conjugate  Green  function $K^{*}$ are  shown  by  
arrows  
$\rightarrow$ and $\leftarrow$, respectively. 
The dashed lines show the transverse  
density  matrices  for  plane  waves  
$\rho_{i}(\ro,\ro')=\exp[i\qb_{i}(\ro-\ro')]$ 
(we assume that for initial particle 
$a$
$\qb_a=0$  at $z=z_i$).
\begin{figure}[h]
\begin{center}
\epsfig{file=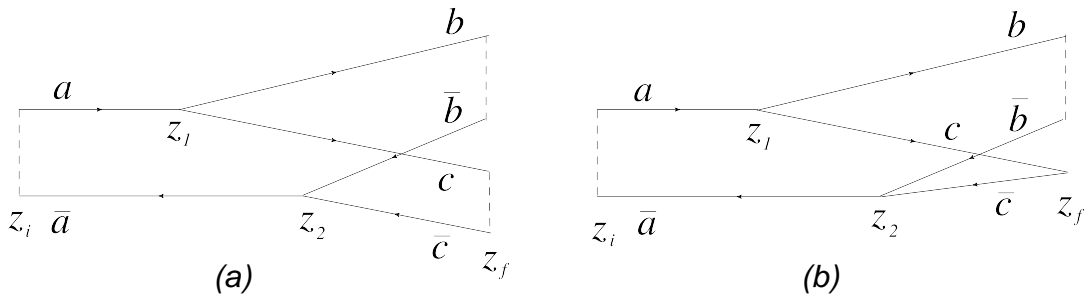,height=3.9cm}
\end{center}
\caption[.]{(a)  Diagram  representation  for  the  spectrum  of
transition $a\to bc$
 in the longitudinal Feynman variable
and the transverse momenta of two final particles in the LCPI
approach.  Dashed  lines  show  the  transverse  density
matrices  of  the  initial  (prior  to  the  interaction  with  the
medium at $z=z_i$) and final (after the interaction with the
medium at $z=z_f$)) particles. (b) The same as in (a) 
for the spectrum   integrated   with   respect   to   
the   transverse momentum of particle $c$.}
\label{fig:pt99-f1}
\end{figure}
\begin{figure}[h]
\begin{center}
\epsfig{file=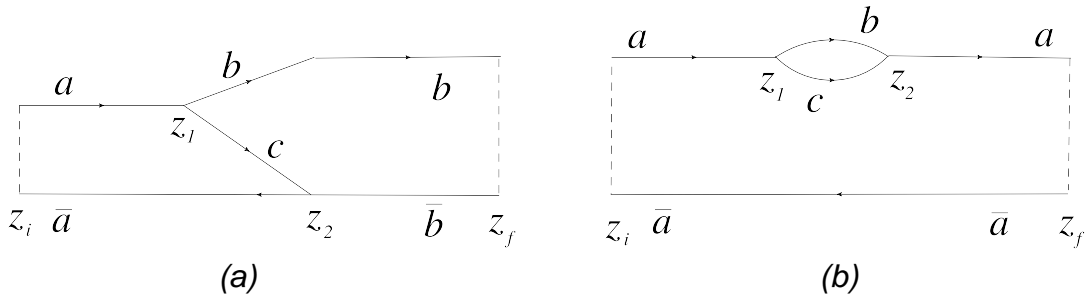,height=3.9cm}
\end{center}
\caption[.]{
(a)  Diagram  representation  for  the  spectrum  of
transition $a\to bc$ integrated with respect to the transverse
momentum  of  particle $c$.
(b)  Diagram  representation  for
the radiative correction to the probability of transition 
$a\to a$ from the virtual process $a\to bc\to a$. There are also 
analogous  diagrams  with  transposition  of  vertices  between  the
upper and lower parts of diagrams (a) and (b).}
\end{figure}

It  should  be  noted  that  the  condition  of  exact
energy conservation in relation (\ref{eq:70-2}) is not necessary for
deriving relation (\ref{eq:80-2}), but it slightly simplifies formulas.
When the potential varies with time, the energy is naturally 
not conserved exactly. It is clear, however, that
if the characteristic time scale for the medium is much
larger than the wavelength of fast particles, the effects
of violation of the energy conservation law for fast particles 
are insignificant for calculation of the probability 
of the process in the leading-order approximation
in  energy.  These  effects  may  give  only  energy-suppressed  
corrections,  the  inclusion  of  which  would
exceed the accuracy of our approximations in the calculation of 
the functions $\phi_{i}$. One can say that for each fast
particle,  what  matters  is  just  the  potential  which  it
“feels”  along  its  trajectory  
$t-z=$const,  and  it  is immaterial  whether  this  potential  
changes  with  time before and after its passage. Physically, this is obvious,
because for a large difference in time/energy scales for
the  medium  and  for  fast  particles,  each  fast  particle
never interacts twice with the same constituent of the
medium.   For   a   time-dependent   potential   of   the
medium,  we  can  also  use  formula  (\ref{eq:80-2}).  In  this  case,  
we must calculate $A^{\mu}$ in the Hamiltonian (\ref{eq:110-2})
 for $\xi=t-z=$const with the same value of 
$\xi$ for the amplitude and
for the complex-conjugate amplitude. When spectrum (\ref{eq:80-2})
is  calculated  without  using  exact  conservation  of
energy (\ref{eq:70-2}), the condition that the functions 
$W(z_1,\ro_1)$ and $W(z_2,\ro_2)$
appear in the expression (\ref{eq:80-2}) for identical values of
$\xi_1$ and $\xi_2$
  appears  after  the  integration  over
the  energy  of  one  of  the  final  particles,  which  gives
$\delta(\xi_1-\xi_2)$. This 
$\delta$-function is then removed by the integration over   
$t_1$,   while  the  integration over
$t_2$ gives just the complete time interval of the
interaction  of  the  incoming  wave  packet  with  the
medium. For the unit time interval, this leads to the formula (\ref{eq:80-2}).

In the LCPI approach, we write all the Green functions  in  
the  evaluation  of  the  transition  probability
described  by  the  diagram  in  Fig.  1a  in  the  Feynman
path integral form \cite{FH}:
\bea
K_i (\ro_2 , z_2 | \ro_1, z_1 ) = 
\int  D \ro \, \exp \left\{ i  \int_{z_1}^{z_2} 
dz \Big[ \frac{\mu_i (d\ro/dz)^2}{2}
- e_iU (\ro , z)\Big] - 
\frac{im^2_i (z_2 - z_1)}{2\mu_i} \right\}\,.
\label{eq:140-2} 
\eea
Here, $U=A^{\mu}v_{\mu}$, where $v_\mu=(1,-d\ro/dz,-1)$
is  the 4-vector of the particle velocity. In the leading order in
energy, one can disregard the transverse component of $v_\mu$
 in calculating the potential\footnote{This corresponds to the disregard 
of $\Ab_\perp$ in the kinetic part of the
Hamiltonian (11). For the static vector potential of the medium,
the omission of $\Ab_\perp$ leads to a loss of the effect of the longitudinal
magnetic field and of the effect of the transverse magnetic field
associated with the derivative of  $\Ab_\perp$ with respect to $z$.
However, for the random vector potential of an amorphous medium, both
these effects are energy-suppressed as compared to the contribution  
from  the term  $A^3$ in the potential $U$
 (and, naturally, from the term $A^0$,
say, in the Coulomb gauge for conventional materials; therefore,
their  inclusion  in  the  approximation  of  the  leading  order  in
energy is meaningless).},
 which gives $U\approx A^{0}-A^{3}$.
After writing all the Green functions in the form of (\ref{eq:140-2}), 
the probability of transition $a\to bc$
 can be represented by a multiple integral over trajectories, 
including the trajectories of particles for the upper and lower
parts  of  the  diagram  in  Fig.  1a.  Integration  is  performed  
over  paths  of  the  particles  in  the  transverse
plane on the light-cone $t-z=$const. For the particle
trajectories  corresponding  to  the  complex-conjugate
Green  functions  at  the  bottom  of  the  diagram  in
Fig. 1a,   the   interaction   with   the   potential   of   the
medium  is  analogous  to  the  interaction  of  antiparticles.  
Therefore,  the  integrand  in  the  functional  integral 
corresponding to Fig. 1a contains the interaction
with  the  medium  in  the  form  of  Wilson's  factors  for
particles  from  the  upper  part  and  antiparticles  from
the  lower  part  (like  in  Fig.  1a,  we  will  denote  the
Green  functions  and  variables  for  lines  with $\leftarrow$
  as belonging to antiparticles).

The main idea of the LCPI method lies in the averaging  
over  the  states  of  the  medium  at  the  level  of  the
integrand prior to evaluation of the functional integrals
in the expression for the transition probability. After this
averaging over the states of the medium, the initial interaction  
of  trajectories  with  a  random  potential  of  the
medium is transformed into the interaction between the
trajectories.  For  the  Abelian  case,  this  interaction  is
described by the effective Lagrangian in the form 
$L_{eff}=in\sigma_X/2$, where $n$ 
 is the number of atoms of the medium
per unit volume and $\sigma_X$
 is the scattering cross section for
the system of particles and antiparticles off a single atom.

Let us consider the spectrum integrated over the 
transverse momentum $\qb_c$. This corresponds to 
the density matrix of the particle $c$ of the form
\beq
\rho_{c}(\ro,\ro')=\frac{1}{(2\pi)^2}
\int d \qb_c\exp{[i(\ro-\ro')\qb_c]}=\delta(\ro-\ro')\,.
\label{eq:150-2}
\eeq
In this case, the diagram in Fig. 1a takes the form of
the diagram in Fig. 1b, which (even prior to the 
averaging  over  the  states  of  the  medium)  can  be  
transformed into the diagram in Fig. 2a without the region
with four trajectories. This transformation is based on
the following identities for the Green functions
\beq
\int \!d\ro_{2}K(\ro_{2},z_{2}|\ro_{1},z_{1})
K^{*}(\ro_{2},z_{2}|\ro_{1}',z_{1})=\delta(\ro_{1}-\ro_{1}')\,,
\label{eq:160-2}
\eeq
\beq
K(\ro_2,z_{2}|\ro_1,z_1)=
\int \!d\ro K(\ro_{2},z_{2}|\ro,z)
K(\ro,z|\ro_{1},z_{1})\,.
\label{eq:170-2}
\eeq
The expression for the spectrum corresponding to the
diagram in Fig. 2a has the form
\bea
\frac{dP}{dx d\qb_{b}}=\frac{2}{(2\pi)^{2}}
\mbox{Re}
\int
d\ro_{bf}d\ro_{\bar{b}f}d\ro_{b2}d\ro_{\bar{b}2}
d\ro_{a1}d\ro_{\bar{a}1}d\ro_{ai}d\ro_{\bar{a}i}
\exp[-i\qb_{b}(\ro_{bf}-\ro_{\bar{b}f})]
\int_{z_{i}}^{z_{f}}dz_{1}\int_{z_{1}}^{z_{f}}dz_{2}
\hat{g} \langle S \rangle\,,
\label{eq:180-2}
\eea
where  subscripts  
$f$, $1$, $2$ and $i$
on  transverse coordinates $\ro$
 indicate that with respect to coordinate
$z$,  they  correspond  to  points
$z_f$, $z_{1,2}$, and $z_i$
  located  as shown in Fig. 2a; like in the initial expression (\ref{eq:80-2}),
$\langle...\rangle$ indicates averaging over the states of the medium, 
and factor $S$  is defined by the relation
\bea
S=S_{b\bar{b}}(\ro_{bf},\ro_{\bar{b}f},z_{f}|
\ro_{b2},\ro_{\bar{b}2},z_{2})
S_{bc\bar{a}}(\ro_{b2},\ro_{c2},\ro_{\bar{a}2},z_{2}|
\ro_{b1},\ro_{c1},\ro_{\bar{a}1},z_{1})
S_{a\bar{a}}(\ro_{a1},\ro_{\bar{a}1},z_{2}|
\ro_{ai},\ro_{\bar{a}i},z_{i})\Big|_{\ro_{c2}=\ro_{\bar{b}2},\ro_{c1}=\ro_{b1}} \,.
\label{eq:190-2}
\eea
The two-particle factors $S_{b\bar{b}}$, $S_{a\bar{a}}$
 are defined by the formula
\beq
S_{i\bar{i}}(\ro_{i2},\ro_{\bar{i}2},z_{2}|
\ro_{i1},\ro_{\bar{i}1},z_{1})=
K_{i}(\ro_{i2},z_{2}|\ro_{i1},z_{1})
K_{\bar{i}}^{*}(\ro_{\bar{i}2},z_{2}|\ro_{\bar{i}1},z_{1})\,,
\label{eq:200-2}
\eeq
while the three-particle factor $S_{bc\bar{a}}$
 for arbitrary positions of the ends of lines
$b$, $c$, and $\bar{a}$ at $z_1$, $z_2$
 is given by
\beq
S_{bc\bar{a}}(\ro_{b2},\ro_{c2},\ro_{\bar{a}2},z_{2}|
\ro_{b1},\ro_{c1},\ro_{\bar{a}1},z_{1})=
K_{b}(\ro_{b2},z_{2}|\ro_{b1},z_{1})
K_{c}(\ro_{c2},z_{2}|\ro_{c1},z_{1})
K_{\bar{a}}^{*}(\ro_{\bar{a}2},z_{2}|\ro_{\bar{a}1},z_{1})\,.
\label{eq:210-2}
\eeq

Typical values of $(z_{2}-z_{1})$
 for the diagrams in Figs. 1a
and  2a  are  determined  by  the  coherence  (formation)
length for transition $a\to bc$, which may considerably
exceed (for relativistic particles) the correlation radius
in  the  amorphous  medium.  Just  in  this  regime  for
transition $a\to bc$
  in  QED  multiple  rescatterings  of
charged particles off atoms of the medium, which are
responsible  for  the  Landau-Pomeranchuk-Migdal
effect,  may  be  important.  In  QCD,  such  a  regime  is
typical for splitting of fast partons in cold and hot QCD
matter. In this regime for an amorphous matter, averaging  over  
the  states  of  the  target  in  the  factor
$S$ in  the expression (\ref{eq:180-2}) can  be  performed  independently  for
individual cofactors; i.e., we can write
\bea
\langle S \rangle =
\langle {S}_{b\bar{b}}\rangle
\langle {S}_{bc\bar{a}}\rangle
\langle{S}_{a\bar{a}}\rangle\,.
\label{eq:220-2}
\eea

Let us first consider evaluating the two-particle
factors $\langle{S}_{i\bar{i}}\rangle$,
each of which is just an evolution operator  of  the  
transverse  density  matrix  for  particle  
$i$. We can write the averaged two-particle factor in the form
of the double path integral
\bea
\langle{S}_{i\bar{i}}\rangle(\ro_{2},\ro_{2}',z_{2}|
\ro_{1},\ro_{1}',z_{1})=
\int  D \ro D\ro' 
\exp \left\{ i  \int_{z_1}^{z_2} 
dz \frac{\mu_i [(d\ro/dz)^2-(d\ro'/dz)^2]}{2}\right\}
{\Phi}_{i\bar{i}}(\{\ro-\ro'\})\,, 
\label{eq:230-2}
\eea
where the functional $\Phi_{i\bar{i}}$ 
 is defined as
\beq
\Phi_{i\bar{i}}(\{\ro-\ro'\})=
\Big\langle
\exp\{-ie_{i}\int_{z_1}^{z_2} dz[U(\ro(z),z)-U(\ro'(z),z)]\}
\Big\rangle\,.
\label{eq:240-2}
\eeq
In the expressions (\ref{eq:230-2}) and (\ref{eq:240-2}), we took into account 
the fact that for a medium invariant to transverse translations, 
the right-hand side of expression (\ref{eq:240-2}) is in fact a
functional of single function $\ta(z)=\ro(z)-\ro'(z)$.
In the case when the length $(z_2-z_1)$ in the two-particle factor 
(\ref{eq:230-2}) is much   larger   than   the   correlation   length   
in   the medium, the functional $\Phi_{i\bar{i}}$
 can formally be written as
\beq
\bar{\Phi}_{i\bar{i}}(\{\ta\})=\exp\left[-\int dz P_i(\ta(z),z)\right]\,,
\label{eq:250-2}
\eeq
where   the   specific   form   of the   function
$P_i(\ta,z)$   
depends on the model of the medium. It can easily be
shown that for the model of the medium in the form
of   randomly   distributed   static   scattering   centers
(atoms), one can obtain
\beq
P_i(\ta(z),z)=\frac{n(z)\sigma_{i\bar{i}}(|\ta(z)|)}{2}\,,
\label{eq:260-2}
\eeq
where $n(z)$ is the local number density of the medium and $\sigma_{i\bar{i}}$
is the total scattering cross section of the $i\bar{i}$   dipole by
an individual atom, which is defined as
\bea
\sigma_{i\bar{i}}(|\ro|)=
2\int d\bb  \Bigg
\{1
-\exp\left[-ie_{i}\int_{-\infty}^{\infty} d\xi\left(\phi(((\bb-\ro)^2+\xi^2)^{1/2})
-\phi((\bb^2+\xi^2)^{1/2}) \right)\right]\Bigg\}\,,
\label{eq:270-2}
\eea
where $\phi(r)$ is  the  potential  of  an  individual  atom.  In
deriving the relations (\ref{eq:250-2}) and (\ref{eq:260-2}) 
from (\ref{eq:240-2}), we took into
account  that  in  the  initial  functional  integral,  the
transverse coordinates of trajectories can be treated as
frozen  on  the  longitudinal  scale  of  the  order  of  the
atomic size.

The fact that ${\Phi}_{i\bar{i}}$  depends only on the relative distance  
between  the  trajectories  allows  one  to  evaluate
the  double  functional  integral  (\ref{eq:230-2}) analytically  
\cite{Zpath87}. The result has the form
\beq
\langle{S}_{i\bar{i}}\rangle(\ro_{2},\ro_{2}',z_{2}|
\ro_{1},\ro_{1}',z_{1})=
K_{i,v}(\ro_{2},z_{2}|\ro_{1},z_{1})
K_{\bar{i},v}^{*}(\ro_{2}',z_{2}|\ro_{1}',z_{1})
{\Phi}_{i\bar{i}}(\{\ta_{l}\})\,,
\label{eq:280-2}
\eeq
where $\ta_{l}$ is a linear function of $z$,
\beq
\ta_{l}(z)=\frac{(\ro_{2}-\ro_{2}')(z-z_1)-(\ro_1-\ro_1')(z-z_2)}{z_2-z_1}\,,
\label{eq:290-2}
\eeq
and $K_{i,v}$ is the free Green function in vacuum,
\beq
K_{i,v} (\ro_2 , z_2 | \ro_1, z_1 ) = 
\frac{\mu_{i}}{2\pi i(z_2-z_1)}
\exp \left[ \frac{i\mu_i (\ro_{2}-\ro_{1})^2}{2(z_{2}-z_{1})}  - 
\frac{im^2_i (z_2 - z_1)}{2\mu_i}\right]\,.
\label{eq:300-2} 
\eeq

The  possibility  of  analytic  evaluation  of  the  functional  
integral  in  the expression  (\ref{eq:230-2})  can  be  expected.
Indeed, the integral in this expression can be written as
the    integral    over  the  center-of-mass    variable
$\Rb=(\ro+\ro')/2$ and $\ta$.    
For the kinetic term in the exponential in (\ref{eq:230-2}), 
we can obtain in these variables
\beq
\int_{z_1}^{z_2} 
dz \mu_i \frac{d\Rb}{dz} \cdot \frac{d\ta}{dz}=
\mu_{i}\left[\left.\Rb \frac{d\ta}{dz}\right|_{z_1}^{z_2}-\int_{z_1}^{z_2}
 dz \Rb\frac{d^{2}\ta}{dz^{2}}\right]\,.
\label{eq:310-2}
\eeq
It can be seen from this expression that the functional
integration  with  respect  to  variable $\Rb$  
  can  be  performed  like  that  for  the  free  Green  functions.  This
integration leads to $\delta(d^{2}\ta/dz^2)$
for each $z$. This $\delta$-function is eliminated by the next 
integration over $\ta$ exactly in the same way as in the free case. Here,
$\delta(d^{2}\ta/dz^2)$ guarantees  that  the  functional $\Phi_{i\bar{i}}$  
  in  the final expression be calculated for function $\tb$, 
 that must be linear in $z$  (since equality $d\ta/dz=$const must hold).
Therefore, the final result must be the product of the
free  Green  functions  by  the  phase  factor  for  a  single
linear trajectory $\ta(z)$.

Let  us  now  consider  the  three-particle  operator $\langle
S_{bc\bar{a}}\rangle$. It is convenient to write the functional integral
$\int D\ro_b D\ro_c D\ro_{\bar{a}}$
  in  new  variables, $\int D\ro D\ro_a D\ro_{\bar{a}}$, where
  $\ro=\ro_b-\ro_c$
 is the relative coordinate for the 
$bc$ system, and $\ro_{a}=x_b\ro_b+x_c\ro_c$
 gives the position of the center of
mass of the $bc$ system. The three-particle phase factor $\Phi_{bc\bar{a}}$
  for  the $bc\bar{a}$ system  prior  to  averaging  over  the states of the
  target is a 
functional of the trajectories in
variables 
$\Rb=(\ro_a+\ro_{\bar{a}})/2$, $\ro_{a\bar{a}}=\ro_a-\ro_{\bar{a}}$ and
$\ro$. Translation  invariance  of  the  system  guarantees  that  the
dependence  on  $\Rb$  in ${\Phi}_{bc\bar{a}}$
  disappears  after  averaging over the states of the matter. Completely analogously to
the  case  of  the  two-particle  operator,  this  allows  us  to
perform  analytic  integration $\int D\ro_a D\ro_{\bar{a}}=\int D \Rb D\ro_{a\bar{a}}$. 
After this, the three-particle factor can be written as
\bea
\langle{S}_{bc\bar{a}}\rangle
(\ro_{b2},\ro_{c2},\ro_{\bar{a}2},z_{2}|\ro_{b1},\ro_{c1},\ro_{\bar{a}1},z_{1})=
K_{a,v}(\ro_{a2},z_{2}|\ro_{a1},z_{1})
K_{\bar{a},v}^{*}(\ro_{\bar{a}2},z_{2}|\ro_{\bar{a}1},z_{1})
{\cal{K}}(\ro_{2},z_{2}|\ro_{1},z_{1})
\,,
\label{eq:320-2}
\eea
where $\ro_{ai}=x_b\ro_{bi}+x_c\ro_{ci}$, $\ro_{i}=\ro_{bi}-\ro_{ci}$,
 for $i=1,2$, and
the last factor is a functional integral with respect to $\ro$
of the form
\bea
{\cal{K}}(\ro_2 , z_2 | \ro_1, z_1 ) = 
\int  D \ro \, \exp \left\{ i  \int_{z_1}^{z_2} 
dz  \frac{M(d\ro/dz)^2}{2}
- 
\frac{i(z_2 - z_1)\epsilon^2}{2M} \right\}
{\Phi}_{bc\bar{a}}(\{\ro\},\{\ro_{a\bar{a}}\})\,.
\label{eq:330-2} 
\eea
Here, $M=E_{a}x_bx_c$,
$\epsilon^{2}=m_{b}^{2}x_{c}+m_{c}^{2}x_{b}-m_{a}^{2}x_{b}x_{c}$,
and $\ro_{a\bar{a}}^{l}$ indicates a function linear in 
$z$,
\beq
\ro_{a\bar{a}}^{l}(z)=\frac{\ro_{a\bar{a}}(z_2)(z-z_1)-
\ro_{a\bar{a}}(z_1)
(z-z_2)}{z_2-z_1}\,,
\label{eq:340-2}
\eeq
which is completely analogous to the function $\ta_{l}$ (\ref{eq:290-2}) for
the two-particle 
operator $\langle{S}_{i\bar{i}}\rangle$ (\ref{eq:280-2}). Averaging over
the states of the target is of local nature with a typical
correlation  length  in  longitudinal  variable  
$z$ on  the order of the atomic size. Therefore, the averaged phase
operator $\Phi_{bc\bar{a}}$
 can formally be written as
\beq
\bar{\Phi}_{bc\bar{a}}(\{\ro\},\{\ro_{a\bar{a}}\}) 
=\exp\left[-i\int dz v(z,\ro(z),\ro_{a\bar{a}}(z))\right]\,.
\label{eq:350-2}
\eeq
Like  in  the  case  of  the  two-particle  phase  factor,  the
form of the function   $v(z,\ro,\ro_{a\bar{a}})$ depends on the model of
the medium, but its specific form is not important for
deriving  the  spectrum.  With  allowance  for  the relation (\ref{eq:350-2}), 
we can state that  ${\cal{K}}$
    is the retarded Green function for the Schr\"odinger equation with the Hamiltonian
\beq
\hat{H}=
\frac{\qb^2+\epsilon^2}{2M}
+v(z,\ro,\ro_{a\bar{a}})
=
-\frac{1}{2M}\,
\left(\frac{\partial}{\partial \ro}\right)^{2}
+v(z,\ro,\ro_{a\bar{a}})
 +\frac{1}{L_{f}}\,.\,\,\,
\label{eq:360-2}
\eeq
Here, we have introduced a quantity
\beq
L_{f}=2E_{a}x_{b}x_{c}/\epsilon^{2}\,,
\label{eq:370-2}
\eeq
which  can  be  viewed as  the  formation  length  for  the
$a\to bc$
 transition in the limit of the low density of the
medium \cite{LCPI1}, because it determines typical scale $z_2-z_1$
for the diagrams in Figs. 1a and 2a in this limit.

For a medium in the form of a system of static scattering  
centers,  the  effective  three-particle  potential  in
the phase factor (\ref{eq:350-2}) and in the Hamiltonian (\ref{eq:360-2}) can
be written as
\beq
v(z,\ro,\ro_{a\bar{a}})=-\frac{i\sigma_{bc\bar{a}}(\ro,\ro_{a\bar{a}})n(z)}{2}\,,
\label{eq:380-2}
\eeq
where
$\sigma_{bc\bar{a}}$ is the cross section of scattering from an atom
of the $bc\bar{a}$  three-particle system. The three-particle cross
section  (and  the potential $v$)  depends  on  longitudinal  variable $x_b$, like “mass” 
$M$ in the Hamiltonian (\ref{eq:360-2}). Like in
the formulas  (\ref{eq:360-2})  and  (\ref{eq:380-2}),  we  will  not  specify  below
explicitly this $x$-dependence.

Substituting  the  resultant  formulas  for  two-  and
three-particle  operators  into  expression  (\ref{eq:180-2}),  we  perform
integration  over  transverse  end  coordinates  for 
$z_i$, $z_1$, $z_2$, and $z_f$, passing to the coordinates of the center of
the mass of pairs and to relative coordinates (for example,
$\Rb_{bf}=(\ro_{bf}+\ro_{\bar{b}f})/2$, 
$\ta_{bf}=\ro_{bf}-\ro_{\bar{b}f}$)
\beq
\int
d\ro_{bf}d\ro_{\bar{b}f}d\ro_{b2}d\ro_{\bar{b}2}
d\ro_{a1}d\ro_{\bar{a}1}d\ro_{ai}d\ro_{\bar{a}i}
=
\int
d\Rb_{bf}d\ta_{bf}d\Rb_{b2}d\ta_{b2}
d\Rb_{a1}d\ta_{a1}d\Rb_{ai}d\ta_{ai}\,.
\label{eq:390-2}
\eeq
Integration  with  respect  to  the coordinates $\Rb$ and $\ta$   
  for  $z_i$ and $z_{1,2}$
 can be performed analytically using the formula
\bea
\int d\Rb_{1} 
K_v(\ro_{2},z_{2}|\ro_{1},z_{1})
K^{*}_v(\ro_{2}',z_{2}|\ro_{1}',z_{1})\nonumber\\
=\left(\frac{\mu}{2\pi (z_2-z_1)}\right)^{2}
\int d\Rb_{1}\exp\left[
\frac{i\mu(\ta_2-\ta_1)(\Rb_2-\Rb_1)}{(z_2-z_1)}\right]=\delta(\ta_2-\ta_1)\,,
\label{eq:400-2}
\eea
where $\ta_i=\ro_i-\ro_i'$ and $\Rb_i=(\ro_i+\ro_i')/2$.
After this integration, the trajectory for segments 
$(z_i,z_1)$ and $(z_2,z_f)$
in the phase factors become parallel, the relative distance 
$\ta_{bf}$  for final $b\bar{b}$  pair being connected with the relative distance 
$\ta_{ai}$ for the initial $a\bar{a}$ pair by the relation 
$
\ta_{i}=x_b\ta_{f}\,
$
(we  will  henceforth  denote  by  $\ta_f$ and $\ta_i$  the  final
and initial vectors, respectively). 
On segment $(z_1,z_2)$ the trajectory of the center of mass of the 
$bc$ pair turns out to be parallel to line $\bar{a}$, and vector $\ro_{a\bar{a}}$,
 appearing in
the  potential (\ref{eq:380-2})  equals $\ta_i$.
Assuming  that  the  total area emerging from the integration with respect to
$\Rb_{bf}$
equals unity, we can write the expression (\ref{eq:390-2}) in the form
\bea
\frac{dP}{dx d\qb_{b}}=\frac{2}{(2\pi)^{2}}
\mbox{Re}
\int
d\ta_f\,\exp(-i\qb_{b}\ta_f)
\int_{z_{i}}^{z_{f}}dz_{1}\int_{z_{1}}^{z_{f}}dz_{2}
\hat{g}\Phi_f(\ta_f,z_{2})
{\cal{K}}(\ro_2,z_{2}|\ro_1,z_{1})
\Phi_i(\ta_i,z_{1})\Big|_{\ro_2=\ta_f,\ro_1=0}
\,,
\label{eq:410-2}
\eea
where
\beq
\Phi_i(\ta_i,z_{1})=
\exp\left[-\int_{z_{i}}^{z_{1}}\!dz P_a(\ta_i,z)\right]\,,
\label{eq:420-2}
\eeq
\beq
\Phi_f(\ta_f,z_{2})=
\exp\left[-
\int_{z_{2}}^{z_{f}}\!dz P_b(\ta_f,z)\right]\,.
\label{eq:430-2}
\eeq

One can expect that at any rate, the typical size of
the  interval  $\Delta z=z_2-z_1$
  in  formula  (\ref{eq:410-2})  should  not
exceed  the  formation  length $L_f$  (\ref{eq:370-2}) for  the  
$a\to bc$ transition  in  vacuum.  However,  the  integration  with
respect  to  variable $z_1$ in  formula  (\ref{eq:410-2})  for  a  finite
medium  in  regions  far  away  from  the  target  (i.e.,  for $|z_{1}|\gg L$)  
should  be  performed  carefully.  Indeed,  the Green  function 
${\cal{K}}$  at  a  large  distance  from  the  target coincides with 
the free Green function
\beq
{\cal{K}}_{v}(\ro_2,z_2|\ro_1,z_1)=
\frac{M}{2\pi i(z_2-z_1)}
\exp\left\{i\left[\frac{M(\ro_2-\ro_1)^{2}}{2(z_2-z_1)}-
\frac{(z_2-z_1)\epsilon^2}{2M}\right]\right\}\,.
\label{eq:440-2}
\eeq
For a free Green function, the integral with respect to
$z_2$ can be expressed in terms of the Bessel function $K_0$
\beq
\int_{z_1}^{\infty}\!
dz_2
{\cal{K}}_{v}(\ro_2,z|\ro_1,z_1)=
\frac{M}{i\pi}K_0(|\ro_2-\ro_1|\epsilon)\,.
\label{eq:450-2}
\eeq
The real part of this integral required for our analysis
is zero. However, this vanishing quantity in our case is
multiplied  by  infinity  due  to  the  integration  with
respect to $z_1$ up to infinity. The elimination of indeterminacy 
$0\cdot\infty$ appearing in this case requires accurate calculations 
with an adiabatically switched off interaction
at large $|z|$. We will perform such calculations for 
$\lambda(z)=\lambda\exp(-\delta |z|)$, followed  by  taking 
the limit $\delta\to 0$. The emergence  of  the contributions  from  
$z$-regions  at  large  distances  from  the  target  is  a  
consequence  of  our  dealing with  the  square  of  the  matrix  element.  
For  matrix  element (\ref{eq:70-2}) itself, the contributions from 
very large values of $|z|$  disappear  due  to  oscillations  of  
the  product  of  
wavefunctions.  However,  for  the  squared  matrix  element,
these oscillations in the matrix element and in the complex-conjugate 
matrix element are canceled out, and the contribution from large $|z|$
should be interpreted carefully.

In the integral over $z_2$ in formula (\ref{eq:410-2}), we
perform identical substitution in the integrand
\bea
\Phi_f(\ta_f,z_{2})
{\cal{K}}(\ro_2,z_{2}|\ro_1,z_{1})
\Phi_i(\ta_i,z_{1})\to
\Phi_f(\ta_f,z_{2})[{\cal{K}}(\ro_2,z_{2}|\ro_1,z_{1})
-{\cal{K}}_{v}(\ro_2,z_{2}|\ro_1,z_{1})]
\Phi_i(\ta_i,z_{1})\nonumber\\
+
[\Phi_f(\ta_f,z_{2})-1]
{\cal{K}}_{v}(\ro_2,z_{2}|\ro_1,z_{1})
[\Phi_i(\ta_i,z_{1})-1]\nonumber\\
+[\Phi_f(\ta_f,z_{2})-1]
{\cal{K}}_{v}(\ro_2,z_{2}|\ro_1,z_{1})
+{\cal{K}}_{v}(\ro_2,z_{2}|\ro_1,z_{1})[\Phi_i(\ta_i,z_{1})-1]
+{\cal{K}}_{v}(\ro_2,z_{2}|\ro_1,z_{1})\,.\,\,\,\,\,\,\,\,\,
\label{eq:460-2}
\eea
After  the  substitution  of  this  expression  into  (\ref{eq:410-2}),  the
last term must vanish since the $a\to bc$ transition does
not occur in vacuum. The first and second terms on the
right-hand side of expression (\ref{eq:460-2}) do not contain contributions
from the distant  regions relative to the target.
The nonzero terms containing the integration over $z$ at
large distances from the target and behind it that are important for eliminating
indeterminacy $0\cdot\infty$ are the first two terms in the last line of
(\ref{eq:460-2}). 
In Appendix A it is shown that the contribution from these terms to
the  spectrum  can  be  expressed  in  terms  of  the  light-cone  
wavefunction  $\Psi$  for  the  Fock  component  
$|bc\rangle$ of particle $a$. This contribution is given by
\bea
\frac{1}{(2\pi)^{2}}\int d\ta_f d\ta'_f
\exp (- i\qb_b  \ta_f)
\Psi^{*}(x , \ta'_f-\ta_f)
 \Psi(x , \ta'_f)
\left[ \Phi_f (\ta_f , z_i ) + \Phi_i (\ta_i , z_f ) - 2 \right]\,.
\label{eq:470-2}
\eea
As a result, the final expression for the spectrum in 
$x$ and $\qb_b$ has the form
\bea
\frac{dP}{dxd\qb_{b}}= \frac{2}{(2\pi)^2} {\rm Re}\!\!  \int\!\!  d\ta  
\exp (- i\qb_{b}\ta_f) 
\int^{z_f}_{z_i}\!\!  dz_1\!  \int^{z_f}_{z_1}\!\! dz_2
\hat{g}\Big\{ \Phi_f(\ta_f , z_2) [{\cal{ K}} (\ro_2 , z_2 | \ro_1 , z_1)\nonumber\\
- {\cal{K}}_{v} (\ro_2 , z_2 | 
\ro_1 , z_1)] \Phi_i (\ta_i , z_1 )
 +
 [ \Phi_f (\ta_f , z_2 ) - 1 ]{\cal{ K}}_{v} (\ro_2 , z_2 | \ro , z_1) 
[\Phi_i (\ta_i , z_1 ) - 1 ]\Big\}\Big|_{\ro_2=\ta_f,\ro_1=0}
\nonumber\\
 + \frac{1}{(2\pi)^{2}}\int d\ta_f d\ta'_f
\exp (- i\qb_b  \ta_f)
\Psi^{*}(x , \ta'_f-\ta_f)
 \Psi(x , \ta'_f)
\left[ \Phi_f (\ta_f , z_i ) + \Phi_i (\ta_i , z_f ) - 2 \right] \,.\,\,\,\,\,\,
\label{eq:480-2}
\eea
Then, after  the  integration  over  the  transverse  momentum,
we obtain the spectrum in one Feynman variable $x$
\bea
\frac{dP}{dx}\! =\! 2 {\rm Re} \!\int^{z_f}_{z_i}\! dz_1\! 
\int^{z_f}_{z_1}\!
dz_2 \hat{g} \left[{\cal{K}} (\ro_2 , z_2 | \ro_1 , z_1)\right.
 -\left. {\cal{K}}_{v} 
(\ro_2 , z_2 | \ro_1 , z_1 )\right]{\Big|}_{\ro_1 = \ro_2 = \ta_f=0}\,.
\label{eq:490-2} 
\eea
For transition of the point-like three-particle system $bc\bar{a}$
at $z_1$ to the point-like system at $z_2$, the Green functions in
this expression must be calculated for $\ro_{a\bar{a}}=0$. Therefore,
potential in the Hamiltonian (\ref{eq:360-2}) becomes central in this case.

Let us now consider how these expressions change
for a fast particle produced in the medium. For particle 
$a$ produced in the medium, it is sufficient to use for
$z_i$ the  coordinate  of  the  production  point  of  the  fast particle $a$;
we take $z_i=0$ for this point. In this case,  in formula (\ref{eq:410-2})
 we perform  the  identity  substitution  in  the  integral  over $z_2$
\bea
\Phi_f(\ta_f,z_{2})
{\cal{K}}(\ro_2,z_{2}|\ro_1,z_{1})
\Phi_i(\ta_i,z_{1})\to
\Phi_f(\ta_f,z_{2})[{\cal{K}}(\ro_2,z_{2}|\ro_1,z_{1})
-{\cal{K}}_{v}(\ro_2,z_{2}|\ro_1,z_{1})]
\Phi_i(\ta_i,z_{1})\nonumber\\
+
[\Phi_f(\ta_f,z_{2})-1]
{\cal{K}}_{v}(\ro_2,z_{2}|\ro_1,z_{1})
\Phi_i(\ta_i,z_{1})\nonumber\\
+{\cal{K}}_{v}(\ro_2,z_{2}|\ro_1,z_{1})[\Phi_i(\ta_i,z_{1})-1]
+{\cal{K}}_{v}(\ro_2,z_{2}|\ro_1,z_{1}) \,.
\label{eq:500-2}
\eea
In  this  case,  indeterminacy  $0\cdot\infty$ appears  only  for  the
range of large positive values of 
$z_{1,2}$ and stems from the
last two terms on the right-hand side of expression (\ref{eq:500-2}).
Obviously,  the  very  last  term  must  give  a  conventional
spectrum corresponding to splitting $a\to bc$ in vacuum,
which, for the initial particle
produced in a hard process,
can now differ from zero  (in contrast to the case with
the initial particle impinging on the target from infinity).
The last but one term in the expression (\ref{eq:500-2}) corresponds
to  the  correction  to  the  vacuum  spectrum  from  rescatterings of the
initial particle in the medium. After eliminating the indeterminacy $0\cdot\infty$ for the last two terms in
this expression by the adiabatic switching off of the interaction for 
$z\to \infty$, the entire spectrum can be written as
\bea
\frac{dP}{dxd\qb_{b}}=  \frac{2}{(2\pi)^2} {\rm Re}  \int  d\ta  
\exp (- i\qb_{b}\ta_f) 
\int^{z_f}_{z_i}\!\!  dz_1\!  \int^{z_f}_{z_1}\!\! dz_2
\hat{g}\Big\{ \Phi_f(\ta_f , z_2) [{\cal{ K}} (\ro_2 , z_2 | \ro_1 , z_1)\nonumber\\
- {\cal{K}}_{v} (\ro_2, z_2 |\ro_1 , z_1)] \Phi_i (\ta_i , z_1 )
 +
 [ \Phi_f (\ta_f , z_2 ) - 1 ]{\cal{ K}}_{v} (\ro_2 , z_2 | \ro_1 , z_1) 
\Phi_i (\ta_i , z_1 )\Big\}\Big|_{\ro_2=\ta_f,\ro_1=0}
\nonumber\\
 + \frac{1}{(2\pi)^{2}}\int d\ta_f d\ta'_f
\exp (- i\qb_b  \ta_f)
\Psi^{*}(x , \ta'_f-\ta_f)
 \Psi(x , \ta')
[\Phi_i (\ta_i , z_f )-1]+\frac{dP_v}{dxd\qb_{b}}\,.\,\,\,\,\,
\label{eq:510-2}
\eea
Here, the last term is the purely vacuum spectrum of
the transition $a\to bc$ 
\beq
\frac{dP_v}{dxd\qb_{b}}=
\frac{1}{(2\pi)^{2}}\int d\ta_f d\ta'_f
\exp (- i\qb_b  \ta_f)
\Psi^{*}(x , \ta'_f-\ta_f)
 \Psi(x , \ta'_f)=\frac{|\Psi(x , \qb_b)|^2}{(2\pi)^2}\,,
\label{eq:520-2}
\eeq
where $\Psi(x , \qb_b)$ is  the  light-cone  wavefunction  in  the
momentum representation for the $a\to bc$ transition.

In  calculating  the  radiative  contribution  to  $p_{\perp}$-broadening of 
particle $b$
 for processes with $a=b$, as for
the $q\to qg$  process  that  will  be  considered  here,  it  is
also necessary to calculate the spectrum for the virtual
process  $a\to bc \to a$ corresponding to the diagram in
Fig. 2b. In the virtual diagram in Fig. 2b, intermediate
system $bc$ evolves from the point-like configuration at
$z_1$ to the point-like configuration at $z_2$. For this reason,
the  Green  functions  appear  in  the expressions  (\ref{eq:410-2}) and
(\ref{eq:510-2}) with  arguments $\ro_2=\ro_1=0$. In this case, in the
terms containing wavefunctions, $\Psi^{*}(x ,\ta_f'-\ta_f)$ is transformed to 
$\Psi^{*}(x ,\ta'_f)$. A distinguishing feature of the virtual diagram is also
that $\ta_i=\ta_f$,
while for a real process, we had $\ta_i=x_b\ta_f$. The final expression for
the contribution of 
the intermediate $bc$ state with a certain value of
longitudinal Feynman variable $x=x_b$ to the spectrum of
the final particle $a$ in transverse momentum  $\qb_a'$    for the diagram in
Fig. 2b has the form (we will use symbol “tilde”
for the quantities in the virtual contribution)
\bea
\frac{d\tilde{P}}{dxd\qb_{a}'}=  -\frac{2}{(2\pi)^2} {\rm Re}  \int  d\ta_f  
\exp (- i\qb_{a}'\ta_f) 
\int^{z_f}_{z_i}\!\!  dz_1\!  \int^{z_f}_{z_1}\!\! dz_2
\hat{g}\Big\{ \Phi_f(\ta_f , z_2) [\tilde{\cal{ K}} (\ro_2 , z_2 | \ro_1 , z_1)\nonumber\\
- \tilde{\cal{K}}_{v} (\ro_2 , z_2 |\ro_1 , z_1)] \Phi_i (\ta_i , z_1 )
 +
 [ \Phi_f (\ta_f , z_2 ) - 1 ]\tilde{\cal{ K}}_{v} (\ro_2 , z_2 | \ro_1 , z_1) 
\Phi_i (\ta_i , z_1 )\Big\}\Big|_{\ro_2=\ro_1=0}
\nonumber\\
 - \frac{1}{(2\pi)^{2}}\int d\ta_f d\ta'_f
\exp (- i\qb_a'  \ta_f)
\Psi^{*}(x , \ta'_f)
 \Psi(x , \ta'_f)
[\Phi_i (\ta_i , z_f )-1]-\delta(\qb_a')\frac{dP_v}{dx}\,,\,\,\,\,\,
\label{eq:530-2}
\eea
where
\beq
\frac{dP_v}{dx}=\int d\qb_b\frac{dP_v}{dxd\qb_{b}}=
\int d\ta_f |\Psi(x , \ta_f)|^2
\label{eq:540-2}
\eeq
is the vacuum spectrum for the $a\to bc$ transition 
 in the Feynman variable $x$. Sign reversal as compared to the spectrum of the
 real process is associated with replacement of the product 
$(i\lambda)(i\lambda)^*$ in the diagram in Fig. 2a by $(i\lambda)^2$
 in the diagram in Fig. 2b. It should be noted that
in  the  above  formulas,  we  did  not  indicate  explicitly
the  dependence  of  the Green  functions  on  the vector $\ta_f$,
which is associated with dependence of the potential
energy (\ref{eq:380-2}) on the vector $\ro_{a\bar{a}}$. The fact that
$\ro_{a\bar{a}}=x_b\ta_f$ for a real
process  and  $\ro_{a\bar{a}}=\ta_f$ for  a  virtual  process  will  be
important in further analysis of the radiative contribution to 
$p_\perp$-broadening.

\subsection{Induced transitions
of type $a\to bc$ for real QED and QCD}
Let  us  first  consider  the  generalization  of  the
expressions  of  the  previous  section  for  real  QED.  In
this  case,  the  three-particle $bc\bar{a}$   system  can  contain
only two charged particles; therefore, the three-particle cross 
section can be expressed in terms of the cross
section  for  the  $e^+e^-$–pair. We must also take into
account the spins of particles in the vertex factor. Let
us  consider  the  generalization  of  the  formulas  of  the
previous section for process
$e\to e\gamma $ (i.e., when $a=b=e$ and $c=\gamma$).

The  $\hat{S}$-matrix element for the $e\to e\gamma$
 process can be written as
\beq
\langle e_f\gamma |\hat{S}|e_i\rangle=-ie\int\! dt
d\rb
\bar{\psi}_{f}\gamma^{\mu}A_{\mu}^{*}\psi_{i}\,,
\label{eq:550-2}
\eeq
where $\psi_{i,f}$  are  the  Dirac  wavefunctions  of  the  initial
and  final  electrons  in  an  external  field  and
$A_{\mu}$   is  the $4$-vector of the wavefunction of the emitted photon. It
is convenient to write the spin states of electrons in the
basis of helicity states in the infinite momentum frame
\cite{BKS,B-L}. Analogously to scalar particles, the electron
Dirac wavefunctions and the wavefunction of the photon
can  be  expresses  in  terms  of  slowly  varying  scalar  functions  
satisfying  the  Schr\"odinger  equation  (\ref{eq:100-2}).  The  
$\hat{S}$-matrix element (\ref{eq:550-2}) can be written in terms of 
the scalar wavefunctions $\phi_i$ for the electron and photon in the form
\beq
\langle  e_f\gamma |\hat{S}|e_i\rangle=
-\frac{i2\pi\delta(E_{\gamma}+E_{e_f}-E_{e_i})}{\sqrt{8E_{e_i}E_{\gamma}E_{e_f}}}\int_{z_i}^{z_f}\! dz\int
d\ro e
\phi_{\gamma}^{*}(z,\ro)\phi_{e_f}^{*}(z,\ro)\hat{\Gamma}\phi_{e_i}(z,\ro)\,,
\label{eq:560-2}
\eeq
Here, $\hat{\Gamma}$    is the vertex operator, which is the sum of the
vertex  operators  conserved  and  flipping the electron
helicity
\beq
\hat{\Gamma}=\hat{\Gamma}_{nf}+\hat{\Gamma}_{sf}\,.
\label{eq:570-2}
\eeq 
The component without the spin flip reads
\beq
\hat{\Gamma}_{nf}=-{\frac{1}{\sqrt{x_{f}}}}\left\{\frac{1+x_{f}}{x_{\gamma}}\qb^{*}\eb^{*}
+i2\lambda[\qb^{*}\times \eb^{*}]_{z}\right\}\,,
\label{eq:580-2}
\eeq
where $\lambda$  is  the  electron  helicity  and $\eb$  is  the  photon
polarization vector, and
\beq
\qb=x_{\gamma}\qb_{f}-x_{f}\qb_{\gamma}\,
\label{eq:590-2}
\eeq
is the operator of the relative transverse momentum for
the pair of final particles $e_f\gamma $. Equation (\ref{eq:560-2}) 
was written in the form in which the momentum operators of the final
particles appearing in $\hat{\Gamma}_{nf}$ were acting from right to left.
The spin-flip component of the operator $\hat{\Gamma}$  is given by
\beq
\hat{\Gamma}_{sf}=
-{\frac{m_e x_{\gamma}}{\sqrt{x_{f}}}}
(2\lambda_{i}e_{x}^{*}+ie_{y}^{*})\delta_{-2\lambda_{f},2\lambda_{i}}\,.
\label{eq:600-2}
\eeq

The presence of the vertex operator $\hat{\Gamma}$ in
expression (\ref{eq:560-2}) does  not  change  the  derivation  of  the  
transition probability as compared to the case of scalar particles.
All  expressions  derived  above  for  scalar  particles  also
hold for the $e\to e\gamma $ transition in real QED if the vertex factor 
(\ref{eq:90-2}) is replaced by the operator
\beq
\hat{g}(z_1,z_2)=\hat{g}_{nf}(z_1,z_2)+\hat{g}_{sf}(z_1,z_2)\,,
\label{eq:610-2}
\eeq
\beq
\hat{g}_{k}(z_1,z_2)=\frac{e^{2}}{16\pi E_{e_i}^{2}x_fx_\gamma}
\hat{V}_{k}(z_1,z_2)\,,
\label{eq:620-2}
\eeq 
where the operator $\hat{V}_{i}$   for the spectrum summed over helicities 
of the final particles and averaged over the helicities
of the initial electron is given by
\beq
\hat{V}_{k}(z_1,z_2)=\frac{1}{2}\sum_{\lambda_{\gamma},\lambda_{i},\lambda_{f}}  
\hat{\Gamma}_{k}(z_1)\hat{\Gamma}_{k}^{*}(z_2)\,.
\label{eq:630-2}
\eeq
Here, arguments $z_{1,2}$ indicate the point at which the operator
$\hat{\Gamma}_{i}$ is acting. Using the expression 
$\sum_{\lambda_{\gamma}}e_{i}(\lambda_{\gamma})e_{j}^{*}(\lambda_{\gamma})=
\delta_{ij}$, from the relations (\ref{eq:580-2}) and
(\ref{eq:600-2})--(\ref{eq:630-2}), one can easily obtain the following 
expressions for the components of $\hat{g}$ for the 
$e\to e\gamma $ process
\beq
\hat{g}_{nf}(z_1,z_2)=\frac{\alpha[1+(1-x_\gamma)^2]}{2x_\gamma
  M^{2}}\qb(z_2)\qb^{*}(z_1)=
\frac{\alpha[1-(1-x_\gamma)^2]}{2x_\gamma
  M^{2}}
\frac{\partial}{\partial \ro_2}\cdot
\frac{\partial}{\partial \ro_1}
\,,
\label{eq:640-2}
\eeq 
\beq
\hat{g}_{sf}(z_1,z_2)=\frac{\alpha m_e^{2}x_\gamma}{2E_{e_i}^{2}(1-x_\gamma)^{2}}\,.
\label{eq:650-2}
\eeq 
In  the  spin  operator  $\hat{g}_{nf}$, after  its  substitution  into
formula (\ref{eq:410-2}), the operators
$\qb(z_i)=-i\partial/\partial\ro_i$
 act on the Green functions for a constant position of the center of mass
of the $bc$ pair. The fact that the operator $\hat{g}_{nf}$ 
 is written in the  form  in  which  the  momentum  operator $\qb(z_2)$  is
acting on the Green function ${\cal{K}}$  (or its vacuum analog),
that  describes  the  intrinsic  dynamics  in  coordinate
$\ro_b-\ro_c=\ro_{f}-\ro_{\gamma}$, may seem strange because initially the
vertex operator in formula (\ref{eq:410-2}) at point $z_2$
 in the lower parts of the diagrams in Fig. 1 acts on the wavefunctions  
of  the final  particles  for  the  complex-conjugate
amplitude. For particles  with  spin,  the diagram in Fig. 1b, 
which corresponds to   the  spectrum   integrated   over   the 
transverse momentum of particle $c$, can also be transformed into the
diagram  of  Fig.  2a.  After this, the momentum operator for $\bar{c}$
in the lower part of the diagram in Fig. 2a now acts on the end of line 
$c$ at the point $z=z_2$, but the momentum operator for $\bar{b}$ continues  
acting on  the  end  of  line $\bar{b}$ at $z=z_2$. However,
using the fact that after averaging over the states of the
medium,  the  averaged  phase  factor  in  the  functional
integral for domain $z>z_2$ depends only on the relative distances  
between  the  trajectories  of  the final  particles,  we
can  transfer  the  differential  operator  of  transverse
momentum  from  the line $\bar{b}$  to  the trajectory  
$b$ in  the  upper part of the diagram, where it approaches the point
$z=z_2$ from  the  left,  by  shifting  the  integration  variables  (in
this  case,  differentiation  does  not  affect  line  
$b$ on the right  of  $z_2$). This  operation  leads  to  
formula  (\ref{eq:640-2}), where momentum operators for 
$z_1$ and $z_2$ are acting on
the Green functions for the three-particle system $bc\bar{a}$ 
at the initial and final points.

Let us now turn to the QCD case. We will treat the QGP  as  a  system  of  static  Debye
screened  color  centers \cite{GW}.  Since  the  exchange  of
$t$-channel gluons between fast partons leads to a change
in their color states as well as color states of the scattering
centers,  the  calculation  of  the induced  splitting  of  partons
in QCD appears at first glance as a more complicated problem  than  
in  the  Abelian  case.  However,  if  at  the amplitude level for the 
$a\to bc$ transition for each center we  account  for  only  one-gluon  
and  color  singlet  two-gluon  exchanges,  then,  the  spectrum  
integrated  over one of the transverse momenta in the two-gluon exchange
approximation for each scattering center is calculated
analogously to the Abelian case. Indeed, the fact that
color generators of any parton 
$p$ and its antipartner $\bar{p}$ are  connected  by  the 
relation  (-$T^{\alpha}_p)^*=T^{\alpha}_{\bar{p}}$,
allows  us  to interpret  the  interaction  of  fast  partons  for  the  lower
part of the diagram in Fig. 1a as the interaction of antipartons. 
Like in the Abelian case, after averaging over the states of the medium 
and summation over all final color states of the medium, there appears 
the interaction of trajectories of fast partons, which is described
by the diffraction operator of a system of partons and
antipartons.  The  difference  between  QED  and  QCD
lies in the fact that for the four-particle part of the diagram in 
Fig. 1a for $z>z_2$, the problem becomes multichannel, because there 
are several color-singlet states for four partons. However, 
for the spectrum integrated with respect of one
of the transverse momenta, which, like in the Abelian
case,  is  described  by  the  diagram  in  Fig.  2a, 
we are dealing with a one-channel problem because there is only 
one color-singlet  state  in  the  intermediate  two-  and  three-particle
states\footnote{At first glance it might seem that two singlet states are possible for 
the $g\to gg$ process  for  the  three-particle  region  
because  there  are  two  singlet color  states  for  three  gluons,  
viz.,  antisymmetric  $\propto f_{\alpha\beta\gamma}$ and
symmetric $\propto d_{\alpha\beta\gamma}$ states.  However,  in  the  case  of  
the $g\to gg$ splitting,  the  system  of  three  partons  in  the  
diagram  in  Fig.  2a
can be only in the antisymmetric color state since after the 
$g\to gg$ transition at $z=z_1$, two gluons are in the 
antisymmetric octet color  state,  and  subsequent  
$t$-channel  gluon  exchanges  cannot change the symmetry of the 
three-gluon color wavefunction.}.
In  this case,  the  diffraction  operator  is  just  the  cross  section
for the corresponding system. As a result, the expression  for  
the  spectrum  has  the  form  analogous  to  the Abelian  case.  
Only  the  expressions  for  the  cross  section  and  for  
the  vertex  factor  change.  We  give here  these
expressions for the $q\to  qg$, process (i.e., $a=b=q$ и $c=g$).
The  main  contribution  to  the  emission  of  a
gluon  by  a  quark  comes  from  the  no  quark  spin  flip
transition. Disregarding the quark spin flip contribution, one can 
write the vertex factor in the form
\beq
\hat{g}(z_1,z_2)=\frac{\alpha_sP_{qq}(x_q)}{2M^2}
\qb(z_2)\qb^{*}(z_1)=\frac{\alpha_sP_{qq}(x_q)}{2M^2}
\frac{\partial}{\partial \ro_2}\cdot
\frac{\partial}{\partial \ro_1}
\,,
\label{eq:660-2}
\eeq
where $P_{qq}$ is the standard splitting function for process
$q\to q$. In the general case of process 
$a\to bc$, one must use  in  the expression (\ref{eq:660-2})  
splitting  function  $P_{ba}(x_b)$. For the  two-gluon  exchange,  
the three-particle  cross  section $\sigma_{bc\bar{a}}=\sigma_{qg\bar{q}}$ 
can  be  expressed  in  terms  of  the  dipole cross section 
$\sigma_{q\bar{q}}$ \cite{NZ_SIGMA3}
\beq
\sigma_{qg\bar{q}}(\ro,\Rb)=
\frac{9}{8}[\sigma_{q\bar{q}}(|\ro|)+
\sigma_{q\bar{q}}(|\Rb-x_b\ro|)]
-\frac{1}{8}\sigma_{q\bar{q}}(|\Rb+x_c\ro|)\,,
\label{eq:670-2}
\eeq
where $\ro=\ro_{b}-\ro_c$ and $\Rb=x_c\ro_b+x_b\ro_c-\ro_{\bar{a}}$.
In  the approximation  of  static  Debye-screened  scattering
centers \cite{GW},  the  dipole  cross  section  for  a color-singlet
$q\bar{q}$ pair has the form
\beq
\sigma_{q\bar{q}}(\rho)=C_{F}C_{R}\int d\qb \alpha_{s}^{2}(\qb^2)
\frac{[1-\exp(i\qb\ro)]}{(\qb^{2}+m_{D}^{2})^{2}}\,,
\label{eq:680-2}
\eeq
where $m_D$ is the Debye mass and $C_F=4/3$ and $C_R$ are the Casimir 
color operators of the quark and of the QGP constituent.   
It   should   be   noted   that   the   scheme
described above in QED makes it possible to take into
account exchanges with any number of 
$t$-channel photons \cite{SLAC1} (for this purpose, it is sufficient 
to calculate the dipole cross section in the eikonal approximation
using  formula  (\ref{eq:270-2})),  while  in  QCD,  our  scheme  works
only  in  the  approximation  of  two-gluon $t$-channel 
exchanges\footnote{In analysis of induced transitions 
$a\to bc$ in QCD in the literature,  the  interaction  of  parton  
trajectories  for  the  diagram  in Fig. 1a is often described in terms of 
Wilson's factors. This may produce  impression  that  the  pattern  
with  a  color-singlet  parton-antiparton system interacting with the 
medium is valid even for  nonperturbative  fluctuations  of  
color  fields  of  the  medium.
However,  there  are  no  grounds  for  this  conclusion,  because  in
the  nonperturbative  situation,  the vector  potentials  in  the  Wilson
lines  for  the  amplitude  and  for  the  complex-conjugate  amplitude 
can be different. Even in perturbation theory at the level of
exchange  of  three  gluons,  the  calculation  of  the  probability  of
the $a\to bc$ transition  cannot be reduced to the problem of passage
of  a  fictitious  parton-antiparton  system  through  the  medium
like in the diagram in Fig. 1a.}.

The  expressions  for  the  static  model  of  the  QGP
can  be  generalized  \cite{AZ}  to  the  case  of  the  dynamic
description  of  the  QGP  in  the  thermal field  treatment  in  the
hard thermal loop (HTL) approximation that was used
in \cite{AMY}. In this case, potential (\ref{eq:380-2}) can be expressed in
terms of the gluon polarization tensor. However, this is
not  expedient,  because  there  are  no  grounds  for  the
applicability  of  the  HTL  scheme  for  the  RHIC  and
LHC conditions. Moreover, it can be shown \cite{RAA13} that
the  HTL  scheme  leads  to  incorrect  normalization  of
the three-particle  potential (\ref{eq:380-2})  for  small-size  parton  states
that  are  important  for  JQ  for  the  RHIC  and  LHC
energies.

\section{CALCULATION  OF  THE  RADIATIVE  
CONTRIBUTION  TO  $p_\perp$-BROADENING
OF  FAST  PARTONS}
We  consider  $p_{\perp}$-broadening  for  a fast  quark  in  a
QGP of finite size $L$ with a uniform density. 
The radiative contribution to $p_\perp$-broadening is connected in this case with the 
$q\to qg$ transition (i.e., $a=b=1$ and $c=g$ in the notation used
in Section 2). We assume that the initial quark is produced with energy 
$E$ at $z=0$.

Let us first consider the conventional nonradiative
$p_\perp$-broadening of a fast quark due to multiple scattering
in  the  medium.  Disregarding  radiative  processes,  we
can  write  the  quark  distribution  over  the  transverse
momentum after its propagation in the medium from
$z_1$ to $z_2$ in terms of the evolution operator of the transverse 
quark density matrix in the form
\beq
\frac{dP}{d\pt}=\int d\Rb_2 d\ta_2 d\ta_1\exp{(-i\pt\ta_2)} 
\langle{S}_{q\bar{q}}\rangle(\ro_{2},\ro_{2}',z_{2}|
\ro_{1},\ro_{1}',z_{1})\,,
\label{eq:690-3}
\eeq
where $\ta_i=\ro_i-\ro_i'$ and $\Rb_2=(\ro_2+\ro_2')/2$.
Using the relation (\ref{eq:280-2}),  from (\ref{eq:690-3}) 
one can easily obtain 
\beq
\frac{dP}{d\pt}=\int d\ta_2\exp{(-i\pt\ta_2)} 
\exp{\left[-\frac{\sigma_{q\bar{q}}(|\ta_2|)Ln}{2}\right]}\,,
\label{eq:700-3}
\eeq
where $L=z_2-z_1$ is the path length in the medium. It
should be noted that although evolution operator (\ref{eq:280-2})
of the density matrix includes the transverse motion of
particles,  the expression  (\ref{eq:700-3})  coincides  with  the  
result  of calculation of $dP/d\pt$ in the eikonal approximation in
which trajectories of particles are assumed to be rectilinear.  
In  QED,  this  fact  was  discovered  in  the  path
integral  method  in \cite{Zpath87}.  We  will  use  the  quadratic
parameterization of the dipole cross section
\beq
\sigma_{q\bar{q}}(\rho)=C\rho^2\,.
\label{eq:710-3}
\eeq 
In   this   approximation,   expression (\ref{eq:700-3})   gives   the
Gaussian distribution
\beq
\frac{dP}{d\pt}=\frac{1}{\pi \langle \pt^2\rangle_0 } 
\exp{\left[-\frac{\pt^2}{\langle \pt^2\rangle_0}\right]}\,,
\label{eq:720-3}
\eeq
where
\beq
\langle \pt^2\rangle_0=2LCn\,.
\label{eq:730-3}
\eeq
This  value  of  the  nonradiative  contribution  to
$\langle p_{\perp}^2\rangle$, with the transport coefficient $\hat{q}=2Cn$
 introduced in \cite{BDMPS1}, corresponds to Eq. (\ref{eq:10-1}). 
The quadratic dipole cross section
approximation  does  not  include  the  Coulomb  logarithmic 
effects in the $\rho$-dependence of the dipole cross
section for $\rho\ll 1/m_D$  and its flattening for
$\rho\gsim 1/m_D$ in the calculation of $\sigma_{q\bar{q}}$ based on 
the two-gluon expression (\ref{eq:680-2}). The  logarithmic  deviation  
from  the  quadratic dependence for small $\rho$ 
 leads to energy dependence of $\langle p_{\perp}^2\rangle$.
If a realistic dipole cross section is used, the value
of $C$ increases  slowly  upon  a  decrease  of $\rho$
  for  small values of $\rho$. In this case, $\langle p_{\perp}^2\rangle$ 
 is given by $2nLC(\rho_{min})$, where $\rho_{min}\sim 1/p_{\perp max}$.
For a quark with energy $E$ in a QGP at temperature 
$T$, we have  $p^2_{\perp max}\sim 3 E T$. The effective
energy-dependent  transport  coefficient $\hat{q}_{eff}=2nC(\rho_{min})$  
can also be written in terms of the differential cross section
$d\sigma/d p_\perp^2$  
 of quark scattering from the constituent of the
medium \cite{BDMPS2,Baier_q,JET_q}
\beq 
\hat{q}=n\int_{0}^{p_{\perp max}^2} dp_\perp^2 p_\perp^2\frac{d\sigma}{d
  p_\perp^2}\,.
\label{eq:740-3}
\eeq

Let us now analyze the radiative contribution to $p_\perp$-broadening. 
We take into account only the one-gluon
emission. In this approximation, initial fast quark 
$q$ at a large distance from the production point may turn
out to be in the one-particle quark state or in the two-parton state $qg$.
It is important that the probability of
formation of the final state $qg$ includes both the conventional 
vacuum splitting and the induced splitting $q\to qg$.
For preserving the total probability, we must take into
account the decrease in the probability of production
of one quark due to possible formation of the two-particle  system.  
This  decrease  in  the  weight  of  the  one-parton  state  
is  described  by  the  radiative  correction
from the virtual process $q\to qg \to q$. We disregard the collisional  
parton energy  loss,  which  is  relatively
small \cite{Z_coll,Gale_coll}. In this approximation, the total energy
of the two-parton state and the energy of the one-parton  state  
are  identical  after  the  passage  through  the
medium. However, the medium may change the relative  weight  of  
the  one-parton  and  two-parton  states.
The  transverse  momentum  distribution  for  partons
also  changes.  We  are  interested  in  the  effect  of  the
medium on the transverse momentum distribution for
the final quark, which is integrated over its
energy.  The  energy  of  the  final  quark  for  the  virtual
contribution  remains  unchanged;  however,  we  must
take  into  account  the  fact  that  rescatterings  in  the
medium for the intermediate two-particle state differ
from  rescatterings  of  a  single  quark.  Since  the  typical
energy  of  the  emitted  gluon  is  much  lower  than  the
quark energy, the ratio of the transverse momentum of
the  final  quark,  which  is  acquired  in  the  medium,  to
the energy can be viewed as the angle of deflection of
the jet due to interaction with the QGP. Therefore,
we can state that our model describes the 
$p_\perp$-broadening in the QGP of the entire jet. In this model, which
exactly  corresponds  to  the  formulation  proposed  in
\cite{Mueller_pt,Blaizot_pt},  the quantity  $\langle p_\perp^2\rangle_{rad}$,
  associated  with  the  interaction with the medium can be written as
\beq
\langle p_\perp^2\rangle_{rad}=\int dx d\pt \pt^2
\left[\frac{dP}{dxd\pt}+\frac{d \tilde{P}}{dxd\pt}\right]\,,
\label{eq:750-3}
\eeq
where $\frac{d P}{dxd\pt}$   is  the  induced  contribution  (i.e.,
without purely vacuum contribution) to the distribution  
in  the Feynman  variable  $x=x_q$  and  the  transverse
momentum of the quark for real process $q\to qg$, and
$\frac{d\tilde{P}}{dxd\pt}$ is the induced contribution to the distribution 
for the virtual process $q\to qg \to q$. In formula (\ref{eq:750-3}),
the meanings of longitudinal variable $x$ for the real and virtual  
contributions  are  different.  For  the  real  process, 
$x$ corresponds to the final quark, while for the virtual process, 
$x$ is determined by the Feynman variable of the quark in the intermediate 
$qg$ state. The variable $p_\perp$ in formula (\ref{eq:750-3}) for 
the real and virtual contributions corresponds to the final quarks. 
By virtue of the energy conservation,  formula (\ref{eq:750-3}) 
can  also  be  written  in  terms  of  the Feynman  variable  for  the  gluon,  
$x_g=E_g/E$,  which  is connected with 
$x_q$ by the relation $x_q+x_g=1$.

In  calculating  the  radiative  contribution  to  
$\langle p_\perp^2\rangle$ based on formula (\ref{eq:750-3}), 
we can avoid the evaluation of the $p_\perp$
  distributions  themselves.  Indeed,  it  can  be  seen
from  the general  expression (\ref{eq:510-2}),  
that  the  induced  spectrum in the transverse momentum for 
the real process for a given $x$ can be written as
\beq
\frac{dP}{dx d\pt}=\frac{1}{(2\pi)^{2}}
\int\!\!
d\ta_f\,\exp(-i\pt\ta_f)F(\ta_f)\,.
\label{eq:760-3}
\eeq
The spectrum (\ref{eq:530-2}) for the virtual process can be written 
in the same form after replacing $F$ by $\tilde{F}$.
It can easily be seen from relations (\ref{eq:750-3}) and 
(\ref{eq:760-3}) that $\langle p_\perp^2\rangle_{rad}$ 
can be expressed in terms of the Laplacian of function
$F+\tilde{F}$ with respect to $\ta_f$ 
at $\ta_f=0$ as
\beq
\langle p_\perp^2\rangle_{rad}=
-\int dx \left.[\nabla^2F(\ta_f)+\nabla^2\tilde{F}(\ta_f)]
\right|_{\ta_f=0}\,.
\label{eq:770-3}
\eeq

Using (\ref{eq:510-2}) for the function 
$F$ of the real process, we can obtain
\beq
F(\ta_f)=F_1(\ta_f)+F_2(\ta_f)\,,
\label{eq:780-3}
\eeq
where  
\bea
F_1(\ta_f)=
2{\rm Re}\!\!  \int^{z_f}_{z_i}\!\!  dz_1\!  \int^{z_f}_{z_1}\!\! dz_2
\Big\{ \Phi_f(\ta_f , z_2) \hat{g}[{\cal{ K}} (\ro_2 , z_2 | \ro_1 , z_1)
- {\cal{K}}_{v} (\ro_2 , z_2 | 
\ro_1 , z_1)] \Phi_i (x\ta_f , z_1 )\nonumber\\
 +
 [ \Phi_f (\ta_f , z_2 ) - 1 ]{\cal{ K}}_{v} (\ro_2 , z_2 | \ro_1 , z_1) 
\Phi_i (x\ta , z_1 )\Big\}\Big|_{\ro_2=\ta_f,\,\ro_1=0}\,,
\label{eq:790-3}
\eea
\beq
F_2(\ta_f)=\int d\ta'_f
\Psi^{*}(x , \ta'_f-\ta_f)
 \Psi(x , \ta')
[\Phi_i (x\ta_f , z_f )-1]\,.
\label{eq:800-3}
\eeq
In (\ref{eq:790-3}) and (\ref{eq:800-3}), we account for the fact that
$\ta_i=x\ta_f$ in (\ref{eq:510-2}). Using the relations 
(\ref{eq:790-3}) and (\ref{eq:800-3}), we obtain the
following  expression  for  the  required  Laplacian  with
respect to $\ta_f$ of $F_{1,2}$ at $\ta_f=0$ 
\bea
\nabla^2F_1(\ta_f)\Big|_{\ta_f=0}
=
2{\rm Re}\!\!  \int^{L}_{0}\!\!  dz_1\!  \int^{\infty}_{z_1}\!\! dz_2
\Big\{ \nabla^2\Phi_f(\ta , z_2) \hat{g}[{\cal{ K}} (\ro_2 , z_2 | \ro_1 , z_1)
- {\cal{K}}_{v} (\ro_2 , z_2 | 
\ro_1 , z_1)]
\nonumber\\+
\nabla^2\hat{g}[{\cal{ K}} (\ro_2 , z_2 | \ro_1 , z_1)
- {\cal{K}}_{v} (\ro_2 , z_2 | 
\ro_1 , z_1)]
 +
\hat{g}[{\cal{ K}} (\ro_2 , z_2 | \ro_1 , z_1)
- {\cal{K}}_{v} (\ro_2 , z_2 | 
\ro_1 , z_1)] \nabla^2\Phi_i (x\ta_f , z_1 )\nonumber\\
+
 \nabla^2\Phi_f (\ta_f , z_2 )\hat{g}{\cal{ K}}_{v} (\ro_2 , z_2 | \ro_1 , z_1) 
\Big\}\Big|_{\ro_2=\ta_f,\ro_1=0, \ta_f=0}\,,
\label{eq:810-3}
\eea
\beq
\nabla^2F_2(\ta_f)\Big|_{\ta_f=0}
=\nabla^2\Phi_i (x\ta_f ,L)\Big|_{\ta_f=0}
\int d\ta'_f
 |\Psi(x , \ta'_f)|^2\,.
\label{eq:820-3}
\eeq
It  should  be  noted  that  in  the  evaluation  of  (\ref{eq:810-3}),
the differential operator $\partial/\partial \ro_1\cdot\partial/\partial \ro_2$ 
 in vertex operator $\hat{g}$ is acting on the Green functions at constant
$\ta_f$, while the
differentiation  with  respect  to $\ta_f$  is the last to be performed.

From (\ref{eq:530-2}) it can be seen  that in the case of
the  virtual  process,  the  expressions  for $\tilde{F}_{1,2}$ can  be
obtained  from  (\ref{eq:790-3}) and (\ref{eq:800-3})  by  replacing  the Green
functions  ${\cal{K}}$ and ${\cal{K}}_{v}$ by $\tilde{\cal{K}}$ and $\tilde{\cal{K}}_{v}$, 
which are calculated now for $\ro_2=\ro_1=0$, and by replacing the argument
$x\ta_f$  in the Glauber factor $\Phi_i$ by $\ta_f$.  We must also reverse
the common signs for $\tilde{F}_{1,2}$. After such transformations, we can write
the  expressions  for  the  Laplacian  of  functions  $\tilde{F}_{1,2}$ in
the form
\bea
\nabla^2\tilde{F}_1(\ta_f)\Big|_{\ta_f=0}
=
-2{\rm Re}\!\!  \int^{L}_{0}\!\!  dz_1\!  \int^{\infty}_{z_1}\!\! dz_2
\Big\{ \nabla^2\Phi_f(\ta_f , z_2) \hat{g}
[\tilde{\cal{ K}} (\ro_2 , z_2 | \ro_1 , z_1)
- \tilde{\cal{K}}_{v} (\ro_2 , z_2 | 
\ro_1 , z_1)]
\nonumber\\+
\nabla^2\hat{g}[\tilde{\cal{ K}} (\ro_2 , z_2 | \ro_1 , z_1)
- \tilde{\cal{K}}_{v} (\ro_2 , z_2 | 
\ro_1 , z_1)]
 +
\hat{g}[\tilde{\cal{ K}} (\ro_2 , z_2 | \ro_1 , z_1)
- \tilde{\cal{K}}_{v} (\ro , z_2 | 
\ro_1 , z_1)] \nabla^2\Phi_i (\ta_f , z_1 )\nonumber\\
+
 \nabla^2\Phi_f (\ta_f , z_2 )\hat{g}\tilde{\cal{ K}}_{v} (\ro_2 , z_2 | \ro_1 , z_1) 
\Big\}\Big|_{\ro_2=\ro_1=0,\ta_f=0}\,,
\label{eq:830-3}
\eea
\beq
\nabla^2\tilde{F}_2(\ta_f)\Big|_{\ta_f=0}
=-\nabla^2\Phi_i (\ta_f ,L)\Big|_{\ta_f=0}
\int d\ta'_f
 |\Psi(x , \ta'_f)|^2\,.
\label{eq:840-3}
\eeq
The  rules  of  action  of  the  differential  operators  in
expression (\ref{eq:830-3}) are the same as in (\ref{eq:810-3}). The
integration with respect to $z_1$ in 
$F_1$ (\ref{eq:810-3}) and $\tilde{F}_1$ (\ref{eq:830-3}) is bounded by
the region 
$z_1<L$, since ${\cal K}-{\cal K}_0$, 
$\tilde{\cal K}-\tilde{\cal K}_v$ and $\nabla^2\Phi_f$ vanish at $z_{1}>L$.
The integration over $z$ in all cases
can be performed for a fixed coupling constant. In our
calculations, it was important to use the adiabatically
switching off coupling constant only in the derivation
of expressions for the terms $F_2$ and $\tilde{F}_2$
via the wavefunction of the two-parton state.

For $\ta_f=0$, the following equality holds for the Green
functions appearing in the expressions for $F_1$ and $\tilde{F}_1$
\bea
\hat{g}
[{\cal K}(\ro_2,z_2|\ro_1,z_1)-{\cal K}_v(\ro_2,z_2|\ro_1,z_1)]\Big|_{\ro_2=\ta_f,\ro_1=0,\ta_f=0}=
\hat{g}[\tilde{{\cal K}}(\ro_2,z_2|\ro_1,z_1)
-\tilde{{\cal K}}_v
(\ro_2,z_2|\ro_1,z_1)]\Big|_{\ro_2=\ro_1=0,\ta_f=0}\,.
\label{eq:850-3}
\eea
Considering  that  the  Glauber  factor $\Phi_f$  for  $F_1$ and  
$\tilde{F}_1$ appears  with  the  same  argument
$\ta_f$, from (\ref{eq:810-3}) and (\ref{eq:830-3})  
we  can see  that  the  terms  with  the Laplacian of $\Phi_f$ 
 in the sum $F+\tilde{F}$ are canceled out exactly. However, 
this property does not hold for the terms containing 
$\nabla^2\Phi_i$  because the quantity $\Phi_i$
 appears in $F_1$ and $\tilde F_1$ with
different arguments. For the same reason, there is no
cancellation  for  $F_2$ and $\tilde F_2$
that  also  contain  different factors $\nabla^2\Phi_i$.
The  terms  in  which  $\nabla^2$ is acting on
the Green  functions  in  $F_1$ and $\tilde F_1$ are  different,  because
the  evaluation  of  the  Laplacian  requires  the  calculation 
of $\hat{g}{\cal{K}}$ and $\hat{g}\tilde{\cal{K}}$
 at nonzero $\ta_f$, for which these functions for the real 
and virtual processes are different.

The  integral  over  the spatial  coordinate  of
the  square  of  the  
$\Psi$-function  in  the expressions  (\ref{eq:820-3}) 
and (\ref{eq:840-3}) gives just the vacuum $x$-spectrum $dP_v/dx$, which
can also be written in terms of the integral over the transverse 
momentum of the double differential spectrum
\beq
\frac{dP_v}{dx}=\int d\pt
\frac{dP_v}{dx d\pt}\,.
\label{eq:860-3}
\eeq
For $q\to qg$ transition, the vacuum spectrum in 
$x$ and the transverse momentum of the quark has the form
\beq
\frac{dP_{v}}{dx d\pt}=\frac{\alpha_sP_{qq}(x)}{2\pi^2}
\frac{\pt^2}{(\pt^2+\epsilon^2)^2}\,,
\label{eq:870-3}
\eeq
where $P_{qq}$ is  the  conventional  splitting  function  for
the $q\to q$ transition. In this case, the integral over
$p_\perp$ in the expression (\ref{eq:860-3}) diverges logarithmically for
large  $p_{\perp}^2$. This  divergence  is  due  to  the  fact  that  we
work  in  the  small-angle  approximation  and  disregard  kinematic  
limits.  In  numerical  calculations,  we regularized this divergence by 
limiting the integration domain  to  
$p_\perp<p_\perp^{max}$ with $p_\perp^{max}=E\mbox{min}(x,(1-x))$.
More  formally,  this  divergence  can  be  regularized  in
the spirit of the Pauli-Villars method by introducing a counterterm 
with replacement of $\epsilon$ by $\epsilon'\sim p_\perp^{max}$.

The  total  contribution  to  $\langle p_\perp^2\rangle_{rad}$  corresponding  
to the sum $F+\tilde{F}$ can be written as the sum of three terms
\beq
\langle p_\perp^2\rangle_{rad}=I_1+I_2+I_3\,,
\label{eq:880-3}
\eeq
where $I_i$  are given by
\bea
I_1=-2\int dx\int_{0}^L dz_1
\int_{0}^{\infty} d\Delta z
\mbox{Re}\Big\{\nabla^2\hat{g}
[{\cal{ K}} (\ro_2 , z_2 | \ro_1 , z_1)-
{\cal{ K}}_v (\ro_2 , z_2 | \ro_1 , z_1)]\Big|_{\ro_2=\ta_f,\ro_1=0,\ta_f=0}
\nonumber\\
-\nabla^2\hat{g}
[\tilde{\cal{ K}} (\ro_2 , z_2 | \ro_1 , z_1)-
\tilde{\cal{ K}}_v (\ro_2 , z_2 | \ro_1 , z_1)]\Big|_{\ro_2=\ro_1=0,\ta_f=0}\Big\}\,,
\label{eq:890-3}
\eea
\bea
I_2=
-2\int dx \int_{0}^L dz_1
\int_{0}^{\infty} d\Delta z
\mbox{Re}\Big\{\hat{g}
[{\cal K}(\ro_2,z_2|\ro_1,z_1)-{\cal K}_v(\ro_2,z_2|\ro_1,z_1)]
\nabla^2\Phi_i(x\ta_f,z_1)\Big|_{\ro_2=\ta_f,\ro_1=0,\ta_f=0}
\nonumber\\
-
\hat{g}[\tilde{{\cal K}}(\ro_2,z_2|\ro_1,z_1)
-\tilde{{\cal K}}_v
(\ro_2,z_2|\ro_1,z_1)]
\nabla^2\Phi_i(\ta_f,z_1)\Big|_{\ro_2=\ro_1=0,\ta_f=0}\Big\}
\nonumber\\
=
-2\langle p_\perp^2\rangle_0
\!\int\!dx f(x)\!\!\int_{0}^L\!\!\! dz_1
\!\frac{z_1}{L}\int_{0}^{\infty}d\Delta z
\mbox{Re}\hat{g}\left[
{\cal K}(\ro_2,z_2|\ro_1,z_1)-{\cal K}_v(\ro_2,z_2|\ro_1,z_1)\right]
\Big|_{\ro_2=\ro_1=0,\ta_f=0}\,,
\label{eq:900-3}
\eea
\bea
I_3=
\int dx
\nabla^2[\Phi_i(\ta_f,L)
-\Phi_i(x\ta_f,L)]\Big|_{\ta_f=0}
\frac{dP_v}{dx}=
-\langle p_\perp^2\rangle_0 \int dx
f(x)\frac{dP_v}{dx}\,,
\label{eq:910-3}
\eea
where $ f(x)=1-x^2$ and $\Delta z=z_2-z_1$. In the expressions
for $I_{2,3}$, we have used relation (\ref{eq:850-3}) and the equalities
\beq
\nabla^2\Phi_i(x\ta_f,z_1)\Big|_{\ta_f=0}=x^2\nabla^2\Phi_i(\ta_f,z_1)\Big|_{\ta_f=0}\,,
\label{eq:920-3}
\eeq
\beq
\nabla^2\Phi_i(\ta_f,z_1)\Big|_{\ta_f=0}=-\langle p_\perp^2\rangle_0z_1/L\,,
\label{eq:930-3}
\eeq
where $\langle p_\perp^2\rangle_0$ corresponds to the nonradiative contribution
(\ref{eq:730-3})  to  $p_\perp$-broadening.  The formulas 
(\ref{eq:880-3})--(\ref{eq:910-3})  are used for numerical calculations of 
$\langle p_\perp^2\rangle_{rad}$. The expressions for the Green functions 
required for calculating $I_{1,2}$ are given in Appendix B.

From  (\ref{eq:910-3}) we can see that $I_3<0$,
which  leads  to  a  decrease  in $\langle p_\perp^2\rangle$.
It will be shown below  that  contribution
$I_2$ for $L=5$ fm is also negative.
This can be explained qualitatively by calculating $I_2$ in
the  approximation  of  the  small  formation  length  for
induced gluon emission as compared to the size of
the medium. In this approximation, we can disregard
the  presence  of  the  boundary  of  the  medium  in  the
integral over $\Delta z$ in (\ref{eq:900-3}). Then, we obtain
\beq
2\mbox{Re}\int_{0}^{\infty}d\Delta z
\hat{g}
[{\cal K}(\ro_2,z_2|\ro_1,z_1)-{\cal K}_v(\ro_2,z_2|\ro_1,z_1)]\Big|_{\ro_2=\ro_1=\ta=0}
\approx 
\frac{dP_{in}}{dxdL}
\,,\,\,\,\,\,
\label{eq:940-3}
\eeq
where $\frac{dP_{in}}{dxdL}$ is  the  $x$-spectrum  of  the induced
gluon emission per unit path length of a quark in the
medium. This leads to the following expression for $I_2$
\beq
I_2\approx
-\frac{\langle p_\perp^2\rangle_0L}{2}
\int\!dx f(x)\frac{dP_{in}}{dxdL}\approx
-\frac{\langle p_\perp^2\rangle_0}{2}
\int\!dx f(x)
\frac{dP_{in}}{dx}\,. 
\label{eq:950-3}
\eeq
Since $\frac{dP_{in}}{dx}>0$, it can be seen that $I_2<0$.
Therefore, the terms proportional to $\nabla^2\Phi_i$
 make a negative contribution to $p_\perp$-broadening.

The integrand in the integral with respect to 
$\Delta z$ in (\ref{eq:890-3}) behaves as $1/\Delta z$  for $\Delta z\to 0$, 
which leads to the logarithmic divergence of $I_1$. Analogously to the
case  of  the  logarithmic  divergence  in  the  integration
over $\pt^2$ for  $dP_v/dx$ in $I_3$,  this  divergence  is  a  consequence  
of  using  the  small-angle  approximation.  The divergence of the 
integral with respect to $\Delta z$ in $I_1$ can also  be  regularized  
by  introducing  the  Pauli-Villars counterterm  
with $\epsilon'\sim p_\perp^{max}$. Such  a  counterterm  will lead to 
cutoff of the integral for $\Delta z\lesssim M/\epsilon'^2$, which 
is equivalent to $\Delta z\lesssim 1/E_g$ for $x_g\ll 1$.
However,  this  procedure could  be  reasonable  only  for  a  
medium  with  the  distance  between  the  constituents  
(and the Debye  radius)  much smaller than $1/M$.
For a real QGP, this inequality does not hold. Therefore, the effect of 
the medium for the real and virtual processes must be small even when 
$\Delta z$ becomes small as compared to the Debye radius. As a
matter of fact, the expressions for the LCPI approach were derived 
under the assumption that the formation length for splitting 
$a\to bc$ for the interaction with an isolated  particle  considerably  
exceeds  the  range  of action of the potential, the role of which is 
played by the  atomic  size  in  QED  and  the  Debye  radius  in  the
QGP. For this reason, for processes in the medium, it is reasonable to 
regularize the integration over $\Delta z$ assuming  that  the  lower  
limit  in the expression (\ref{eq:890-3}) is $\Delta z\sim 1/m_D$ 
(this value is substantially bigger than $1/E_g$ at $E_g\gg m_D$).
This prescription was proposed in  \cite{Mueller_pt}
for  calculating  the  radiative  contribution  to  $p_\perp$-broadening  
with  a  logarithmic  accuracy.  In  our  formulation,  
the  contribution  considered  in \cite{Mueller_pt}  stems
from the factor $I_1$ (\ref{eq:890-3}). The authors have observed that
the  predominant  contribution  to $\langle p_\perp^2\rangle_{rad}$ comes  
from the   double   logarithmic   integral   
$\int dx_g/x_g\int d\Delta z/\Delta z$, 
which leads exactly to the formula (\ref{eq:20-1}) for $\Delta z_{min}=l_0$.
In \cite{Mueller_pt}, the authors did not account for
the  Glauber  factors $\Phi_{i,f}$ in evaluating 
$\langle p_\perp^2\rangle_{rad}$. For  this  reason,  contributions  
$I_{2,3}$, proportional  to $\nabla^2 \Phi_i$ have been missed. 
As mentioned above, these contributions are negative and lead to a weaker
$p_\perp$-broadening.  It  will  be  seen  from  the  results  of
numerical  calculations  that  the  total  negative  contribution of
$I_2$ and $I_3$ for  the RHIC and LHC conditions is
larger  in  magnitude  than  $I_1$,  and  the  value  of 
$\langle p_\perp^2\rangle_{rad}$ turns out to be negative.

\section{ RESULTS OF NUMERICAL CALCULATIONS}
In  numerical  calculations  of  $\langle p_\perp^2\rangle_{rad}$,  we  
used  the values $m_q=300$ MeV and $m_g=400$ MeV as the main
variant for the quasiparticle masses, which were obtained
from  analysis  of  the  lattice  data  in  the  quasiparticle
model of the QGP  \cite{LH} for temperatures corresponding to 
the RHIC and LHC conditions. With these values  of  masses,  in  
our  previous  works \cite{RAA13,RPP14} on  JQ,  we  successfully  described  
the  RHIC  and  LHC  data  on the nuclear  modification  factor  
$R_{AA}$.  The  results  for  $R_{AA}$ are  not  very  sensitive  to  
the quasiparticle  masses. To   understand   the uncertainties   
associated   with   the choice  of  the parton  masses,  we  also performed  
calculations for masses $m_{q}=150$ MeV and  $m_{g}=200$ MeV.
In this study,  like  in  \cite{Mueller_pt},  calculations  are
 performed  with constant $\hat{q}$ and fixed $\alpha_s$ at the vertex of 
the decay $q\to qg$.
In  \cite{RAA13,RPP14}, $R_{AA}$  was  calculated  in  a  more  realistic
model beyond the oscillator approximation and using
running $\alpha_s$  with  the  Debye  mass  of  the  QGP,  which
was obtained in lattice calculations \cite{Bielefeld_Md}. Also,
the calculations  of  \cite{RAA13,RPP14},  were  performed  accounting  for  
the longitudinal expansion  of  the  QGP  in  the  Bjorken
model \cite{Bjorken}, which leads to dependence of the transport 
coefficient on the proper time, $\hat{q}\propto 1/\tau$.
For  more  reliable  predictions  concerning   
$p_\perp$-broadening  in  the  model  with  fixed  
$\alpha_s$  without  the QGP expansion, we have performed fitting of 
the parameter $\hat{q}$   from the condition of coincidence of 
the quark energy loss $\Delta E$ in  the  formulation  of  this  
study  to  the  results  of  the more realistic model used in \cite{RPP14}. 
For the conditions of central Au+Au collisions at RHIC for $\sqrt{s}=0.2$
TeV, we  obtained  transport  coefficient\footnote{In this study, we use
in all formulas the transport coefficient
 of the quark which is smaller than the gluon transport coefficient by a
factor of $C_F/C_A=4/9$.} $\hat{q}\approx 0.12$ GeV $^3$ for $E=30$ GeV.
For  the  Pb+Pb  collisions  at  LHC  for
$\sqrt{s}=2.76$ TeV, we obtained $\hat{q}\approx 0.14$ GeV$^3$ for $E=100$
GeV. Like  in  the  calculations  performed  in  \cite{Mueller_pt}
we take  $\alpha_s=1/3$ and $L=5$ fm (this value of $L$, approximately 
corresponds to the typical path length of a jet for central collisions).

In the above formulas for $I_{2,3}$, we expressed $\nabla^2\Phi_i$
in terms  of the  nonradiative $\langle p_\perp^2\rangle_{0}$,  which  
is  connected  with the transport coefficient $\hat{q}$  by the 
relation (\ref{eq:10-1}). In the oscillator approximation   that   
is   used   here   for   calculating the Green  functions  appearing  in  
the  expressions  for $I_{1,2}$,  the frequency
$\Omega$  in  the oscillator  Hamiltonian  (\ref{eq:1150}) 
in Appendix  B  also  contains $\hat{q}$ (since $\Omega^2\propto \hat{q}$).
It should be borne  in  mind  that  these  values  of $\hat{q}$ can  differ  
from each other due to the Coulomb effects. Indeed, in its physical meaning, 
the quantity $\hat{q}$ appearing in the calculation of $\nabla^2\Phi_i$
 corresponds to rescatterings in the medium
of the initial quark. Therefore, it is natural to take
$\hat{q}=2nC(\rho\sim 1/p_{\perp max})$,
as  was discussed in  Section  3.  At  the
same time, it is natural to use the transport coefficient
defined  as  $2nC$,  where  $C$
  is  the  ratio  of  exact  dipole
cross section (\ref{eq:680-2}) to $\rho^2$ 
 at $\rho_{eff}$ defined as the characteristic size of the three-parton 
system, for frequency $\Omega$ in the Hamiltonian (\ref{eq:1150}) describing
the Green functions \cite{LCPI1,LCPI_YF98}. For the 
$q\to qg$ transition for the RHIC and LHC conditions,  this  method  gives  
the  value  of $\hat{q}$ close  to $\hat{q}$
that is determined by formula (\ref{eq:740-3}) for the 
energy equal to the  typical  energy $\bar{E}_g$ of  the  emitted  gluon,  
which  is much  lower  than  energy  $E$ of  the  initial  quark  and
weakly depends on the quark energy. For quarks with $E\sim  30-100$ GeV 
for the RHIC and LHC conditions, we  have $\bar{E}_g\sim 3-5$. 
In  this  case,  the  variation  of the  transport  coefficient  with  
energy  in  the  calculation based on formula (\ref{eq:740-3}) 
turns out to be significant. We will  denote  the  transport  
coefficient  of  the  initial quark as $\hat{q}'$, retaining the notation 
$\hat{q}$ for the coefficient of the gluon energy, 
which appears in expression for frequency  in  the  oscillator  
Hamiltonian  (\ref{eq:1150})  for  the three-parton  system.  
Our  calculations  based  on (\ref{eq:740-3}) with running $\alpha_s$  
and with the Debye mass of the QGP predicted by the lattice 
calculations \cite{Bielefeld_Md} give the following value for
$r=\hat{q}'/\hat{q}$
\beq
r\approx 1.94(2.13)
\label{eq:960-4}
\eeq
for quarks with energy $E=30(100)$ GeV for the RHIC(LHC) conditions.

In numerical calculations in (\ref{eq:890-3})--(\ref{eq:910-3})
we integrate over $x$ from $x_{min}=m_q/E$
up to $x_{max}=1-m_g/E$ (recall that we define  $x$ as $x_q$;
in terms of $x_g$, our domain corresponds to the variation
of $x_g$ from $m_g/E$ to $1-m_q/E$). Like in \cite{Mueller_pt}, we regularize 
the $1/\Delta z$ divergence in (\ref{eq:890-3}) by truncating  the  
 integration  at $\Delta z_{min}=1/m$ with $m=300$ MeV.
Our  numeric calculations  give for  the terms  
$I_{1,2,3}$ in formula (\ref{eq:880-3})
\beq
[I_1,I_2,I_3]/\langle p_\perp^2\rangle_{0}\approx
[0.417/r,-0.213,-0.601]\,
\label{eq:970-4}
\eeq
for $E=30$ GeV for the RHIC conditions. Calculations
for   the   LHC   conditions   for the  quark   energy   
$E=100$ GeV give
\beq
[I_1,I_2,I_3]/\langle p_\perp^2\rangle_{0}\approx
[0.823/r,-0.107,-0.908]\,.
\label{eq:980-4}
\eeq

Using the values of the ratio  $\hat{q}'/\hat{q}$ from (\ref{eq:960-4}), we obtain from
relations  (\ref{eq:970-4}) and (\ref{eq:980-4}) the  following  values  for  the
ratios  of  the  radiative  and  nonradiative  contributions
in our versions for RHIC(LHC)
\beq
\!\langle p_\perp^2\rangle_{rad}/
\langle p_\perp^2\rangle_{0}
\approx -0.598 (-0.629)\,,\,\,
r=  1.94(2.13)\,.
\label{eq:990-4}
\eeq
And for $\hat{q}'=\hat{q}$ we obtain
\beq
\!\langle p_\perp^2\rangle_{rad}/
\langle p_\perp^2\rangle_{0}
\approx -0.397(-0.192)\,,\,\,
r=1(1)\,.
\label{eq:1000-4}
\eeq
It can be seen that even in the version disregarding the
difference between $\hat{q}'$ and $\hat{q}$, the radiative contribution
to  the  $p_{\perp}$-broadening  turns  out  to  be  negative  for  the
RHIC and LHC conditions.

As we have said, to investigate the sensitivity of the results to parton
masses,  we  also  performed  calculations  for  half  as
large   parton   masses   ($m_q=150$ MeV and $m_g=200$ MeV). This leads 
to an increase in the magnitudes
of the contributions $I_{1,2,3}$ by $\sim 10-20$\%.  The sensitivity of the
total $\langle p_\perp^2\rangle_{rad}$  to the reduction of masses by
half turns out to be slightly higher (since there exists a
strong compensation between the contribution from $I_1$ and the negative contributions from 
$I_{2,3}$. The values of the total $\langle p_\perp^2\rangle_{rad}$ in
all versions remain negative. For the version with $\hat{q}'>\hat{q}$ 
(\ref{eq:960-4}), 
the absolute value of $\langle p_\perp^2\rangle_{rad}$ 
 increases approximately by a factor of $1.36 (1.4)$ for RHIC(LHC), while
in the version with  $\hat{q}'=\hat{q}$, it increases by a factor of $\sim
1.26(1.5)$ for RHIC(LHC). 
The main negative contribution to $\langle p_\perp^2\rangle_{rad}$ comes from
the term $I_3$. In the above results on the dependence on the parton
masses, the vacuum spectrum appearing in the expression (\ref{eq:910-3}) for 
$I_3$ has been calculated for the quasiparticle parton masses in the QGP.
We also investigated the change in the results in the
case when the vacuum spectrum was calculated for the gluon mass $m_g=800$ MeV.
Approximately such a gluon mass was obtained in \cite{NZ_HERA} 
($m_{g}=750$ MeV) from analysis  of  the  proton  structure  function  $F_2$
for small $x$ within the  dipole BFKL equation. 
The gluon mass obtained in \cite{NZ_HERA} is  in  good
agreement  with  the  natural  infrared cutoff  for  perturbative  gluons,
$m_{g}\sim 1 / R_{c}$, where $R_{c}\approx 0.27$ fm is  the  gluon
correlation  radius  in  the QCD vacuum \cite{shuryak1}. Consequently, the choice of 
$m_g\sim 800$ MeV appears  as  reasonable.  It  should  be  noted  that  the
LCPI formalism permits in principle the use of parton masses  depending  on
the longitudinal  coordinates.  Our calculations with 
$m_g=800$ MeV lead to suppression of $I_3$ by a factor of $0.77(0.83)$ for the
RHIC(LHC) conditions. In this case, ratio $\langle p_\perp^2\rangle_{rad}/\langle p_\perp^2\rangle_{0}$
 remains negative  both  in  the  version  with  $\hat{q}'>\hat{q}$ and  with  $\hat{q}'=\hat{q}$       
(estimates obtained with the large mass $m_g$ for the vacuum  spectrum  are  
naturally  qualitative  since  in  the case of different parton masses in the
QGP and in vacuum,   the   influence   of   the   Ter-Mikaelyan   effect
should also be taken into account \cite{Z_TM}).

Thus, our tests have shown that the prediction concerning the negative 
value of $\langle p_\perp^2\rangle_{rad}$ is quite insensitive
to the parton masses. It should be noted that the sensitivity  of  the induced gluon
emission  to  the  mass  of the light quark is generally low (except for the emission
of hard gluons with $x_g\sim 1$, and the change in the predictions is mainly 
associated with variation of $m_g$.

In the results presented above, we have used fixed $\alpha_s$.
The generalization of calculations to running $\alpha_s$
is a  complicated  problem  which  is  beyond  the  scope  of
this article. At the same time, using the scheme proposed in this study, 
one can easily estimate the effect of running $\alpha_s$ on the predominant 
negative contribution  to  $\langle p_\perp^2\rangle_{rad}$
  from  the term  $I_3$,  which  is  associated  with the  dependence  
on  the  parameterization  of $\alpha_s$ of  the
purely vacuum spectrum $dP_{v}/dx$ in formula (\ref{eq:910-3}).
For calculating $dP_{v}/dx$  with  the  running  coupling  constant, 
it is sufficient in formula (\ref{eq:870-3})  to replace  
the static $\alpha_s$ by the running one. We have used the one-loop
$\alpha_s$  frozen for small momenta at value  $\alpha_{s}^{fr}=0.7$.
This  value  of $\alpha_{s}^{fr}$   for  the  given  parameterization  was
obtained  earlier  from  analysis  of  the structure  functions
for  small  $x$ based on the dipole BFKL equation \cite{NZ_HERA}.
This  value  matches  well  to  the  result  of  analysis  of
heavy  quark  energy  loss  in  vacuum  \cite{DKT}. It  should  be
noted  that  the  method  for  calculating  the factor $\nabla^2\Phi_i$
appearing in formula (\ref{eq:910-3}) is immaterial at all for the ratio
$I_3/\langle p_\perp^2\rangle_{0}$ of interest to our analysis. 
The use of the vacuum  spectrum  with  such  running $\alpha_s$
  leads  to  an increase in the absolute value of $I_3$ by a
factor  of  $\sim 1.45(1.2)$1.45 for  the  RHIC(LHC)  conditions.
The  absolute  value  of  the ratio 
$\langle p_\perp^2\rangle_{rad}/\langle p_\perp^2\rangle_{0}$   increases  in
this   case   by  a factor of $\sim 1.45(1.3)$
for  RHIC(LHC). 

The  large  relative  contribution  from $I_{2,3}$ renders
our  results  for  the  radiative  contribution  to  $p_\perp$-broadening  
radically  differing  from  the  appreciable positive 
radiative correction $\langle p_\perp^2\rangle_{rad}\approx 0.75\hat{q}L$
 predicted in \cite{Mueller_pt}. In the form used in the expressions 
(\ref{eq:990-4}) and (\ref{eq:1000-4}), this corresponds to
$\langle p_\perp^2\rangle_{rad}/\langle p_\perp^2\rangle_{0}\approx 0.75/r$.
This prediction  is  in  qualitative  agreement  with  our  results 
(\ref{eq:970-4}) and (\ref{eq:980-4}) for  the  contribution  
to  $\langle p_\perp^2\rangle_{rad}$ 
  from  single term $I_1$, which can be treated as an analog of the result
obtained  in  \cite{Mueller_pt} (but  with  careful  numerical  
calculation  beyond  the  logarithmic  approximation  and  the
soft gluon approximation)

Note that the inclusion of the terms $I_{2,3}$,
that have been disregarded in \cite{Mueller_pt}, also changes the physical
pattern of the radiative $p_\perp$-broadening. Indeed, the contribution  
from  $I_1$  in  the  approximation  of  small  formation length
$L_f\ll L$ can be viewed qualitatively as a local effect in the 
longitudinal coordinate and can be interpreted  as  a  
renormalization  of  the  transport  coefficient. On the contrary, 
for the contributions of $I_{2,3}$, the longitudinal distances $\sim L$ 
are important. 
For this reason, the effect of the terms $I_{2,3}$ on the $p_\perp$
 broadening cannot be interpreted  as  simple  renormalization  of  the  local
transport coefficient. It is important that for the dominating negative
contribution from $I_3$ the  gluon emission  can occur  in  vacuum.  
This  fact  casts  a  shade  of doubt on the possibility of 
factorization of the effects of interaction with the medium and 
Sudakov's effects in analysis of the azimuthal jet decorrelation in 
$AA$ collisions, as it was done in \cite{Mueller_dijet}.

\section{CONCLUSIONS}
We  have  analyzed  the  radiative $p_\perp$-broadening  of fast  partons  in  a  QGP.  Analysis  has  been  performed
using  the  LCPI  formalism  \cite{LCPI1,LCPI_PT}  in  the  oscillator
approximation. Calculations have been carried out for
a homogeneous QGP of thickness  $L=5$ fm with values  of  the  
transport  coefficient  corresponding  to  the
conditions of central nuclear collisions of Au+Au and
Pb+Pb at RHIC and LHC. It is shown that the contributions  to  
the  radiative  $p_\perp$-broadening  come  from
both real and virtual processes that are local by nature over the 
jet path length with a characteristic longitudinal  size  on  
the  order  of  the formation  length  of  the induced gluon emission, 
as well as from the processes including rescatterings of the 
initial parton over sizes on the order of  the  size  of  the  QGP.  
Processes  of  the  former  type make  a  positive  contribution  to   
$p_\perp$-broadening, while  the  nonlocal  processes  of  the  
latter  type,  conversely,  make  a  negative  contribution  and  reduce  
$p_\perp$-broadening. The processes of the first type were considered  
earlier  in  \cite{Mueller_pt,Blaizot_pt} to logarithmic  accuracy  
in  the  soft  gluon  approximation.  The  contribution from 
the initial parton rescatterings to $p_\perp$-broadening is 
considered for the first time.

Our calculations have shown that for the RHIC and LHC  conditions,  
the  negative  contribution  from  the initial parton rescatterings is so 
large that the total $\langle p_\perp^2\rangle_{rad}$ turns  out  to  
be  negative  and  can exceed in absolute value the half of 
traditional nonradiative contribution $\langle p_\perp^2\rangle_{0}$.
In this case, the total effect of  nonradiative  and  radiative  
mechanisms  on  $p_\perp$-broadening of jets may turn out to 
be quite small. This probably explains a slightly unexpected negative result
of the STAR experiment \cite{STAR1} aimed at the search for
the effect of jet rescatterings in a QGP in Au+Au collisions  at
$\sqrt{s}=0.2$ TeV.  Naturally,  it  is  extremely important to 
generalize the calculations performed in this  study  to  the  case  of  
expanding  QGP  to  draw  a more reliable conclusion.\\

\section*{Acknowledgments}
I am grateful to the chief editor  of JETP  Academician  
A.F. Andreev  for suggestion  to
submit this article for the jubilee issue of the journal devoted
to the centenary of Academician I.M. Khalatnikov.

\section*{Appendix A}
We consider here the elimination of the indeterminacy $0\cdot\infty$ 
emerging from the regions of large $z_{1,2}$ in formula (\ref{eq:410-2}). Let
us calculate the contribution to the spectrum
in $x$ and $\qb_b$ for process $a\to bc$ from the term 
${\cal{K}}_{v}(\ro_2,z_{2}|\ro_1,z_{1})[\Phi_i(\ta_i,z_{1})-1]$
in (\ref{eq:460-2}), for which indeterminacy  $0\cdot\infty$ appears  (we
denote  it  by $dP_{+}/dxd\qb_b$). To resolve the indeterminacy $0\cdot\infty$, the contribution of a
finite  region  in  $z_1$ is  insignificant;  therefore,  we  can write
\bea
\frac{dP_{+}}{dx d\qb_{b}}=\frac{2}{(2\pi)^{2}}
\mbox{Re}
\int
d\ta_f\,\exp(-i\qb_{b}\ta_f)
\int_{0}^{\infty} dz_{1} \int_{z_{1}}^{\infty}dz_{2}
\hat{g}
{\cal{K}}_{v}(\ro_2,z_{2}|\ro_1,z_{1})[\Phi_{i}(\ta_i,\infty)-1]
\Big|_{\ro_2=\ta_f,\ro_1=0}\,.
\label{eq:1010}
\eea
We write the Green function in the form of the Fourier representation
\beq
{\cal{K}}_{v}(\ro_2,z_{2}|\ro_1,z_{1})=\frac{1}{(2\pi)^2}
\int d\qb \exp{[i\qb(\ro_2-\ro_1)]}
\exp{\left[-i(z_2-z_1)\frac{\qb^2+\epsilon^2}{2M}\right]}\,.
\label{eq:1020}
\eeq
We  take  the  interaction  constant  in  the  form
$\lambda(z)=\lambda\exp(-\delta|z|)$, taking  the  limit $\delta \to 0$ in
the final  expressions.
Separating explicitly the exponential $z$-dependence of $\hat{g}$, we obtain
for a fixed $\delta$
\bea
\frac{dP_{+}}{dx d\qb_{b}}=\frac{2\hat{g}}{(2\pi)^{4}}
\mbox{Re}
\int
d\qb J(\qb_b-\qb)
\int
\limits_{0}^{\infty}\!\!dz_{1}
\exp{(-2\delta z_1})
\int\limits_{0}^{\infty}\!\!d\xi
\exp{\left[-\delta \xi -i\xi \frac{\qb^2+\epsilon^2}{2M}\right]}\,,
\label{eq:1030}
\eea
where
\beq
J(\kb)=\int d\ta_f \exp{(-i\kb\ta_f})
[\Phi_{i}(x\ta_f,\infty)-1]\,.
\label{eq:1040}
\eeq 
In this relation, we consider that $\ta_i=x\ta_f$ (we assume that $x=x_b$).
After integration over $z_1$
 and $\xi$ and passing to the limit $\delta\to 0$, we obtain
\bea
\frac{dP_{+}}{dx d\qb_{b}}=\frac{\hat{g}}{(2\pi)^{4}}
\int
d\qb J(\qb_b-\qb)\left(\frac{2M}{\epsilon^2+\qb^2}\right)^2\,.
\label{eq:1050}
\eea
Using  the  noncovariant  perturbation  theory  in  the
infinite  momentum  frame,  one  can  easily  show  that
the wavefunction for two-particle Fock state $|bc\rangle$ in the
$(x,\qb)$-representation for the $a\to bc$ transition reads
\beq
\Psi(x,\qb)=\frac{\lambda \sqrt{x(1-x)}}{2\sqrt{\pi}(\epsilon^2+\qb^2)}\,.
\label{eq:1060}
\eeq 
Here, $\Psi(x,\qb)$ is normalized so that the probability of
the Fock component $bc$  in the physical particle $a$ is
\beq
P(a\to bc)=
\frac{1}{(2\pi)^2}\int dx d\qb |\Psi(x,\qb)|^2\,.
\label{eq:1070}
\eeq 
With  allowance  for  relations  (\ref{eq:1060}) and (\ref{eq:90-2}),  we  can
write (\ref{eq:1050}) in the form
\bea
\frac{dP_{+}}{dx d\qb_{b}}=\frac{1}{(2\pi)^{4}}
\int
d\qb J(\qb_b-\qb)|\Psi(x,\qb)|^2\,.
\label{eq:1080}
\eea
This expression can be written in the coordinate representation as
\bea
\frac{dP_{+}}{dx d\qb_{b}}=\frac{1}{(2\pi)^{2}}
\int
d\ta_f d\ta'_f \exp{(-i\qb_b\ta_f)} \Psi^{*}(x,\ta'_f-\ta_f)\Psi(x,\ta'_f)
[\Phi_i(x\ta_f,\infty)-1]\,.
\label{eq:1090}
\eea

The  contribution  from  the  region  of  negative  
$z_{1,2}$ for (\ref{eq:410-2}) from the term $[\Phi_f(\ta_f,z_{2})-1]{\cal{K}}_{v}(\ro_2,z_{2}|\ro_1,z_{1})$ 
in (\ref{eq:460-2}) can be calculated analogously, and the total  contribution
from  the regions  $z_{1,2}<0$ and $z_{1,2}>0$ leads  to  relation (\ref{eq:470-2}).
At  the  same  time,  the expression (\ref{eq:1090}) gives the contribution to
the spectrum in the situation  with  initial  particle $a$
  produced  at  $z=0$. The application of an analogous method for the last term
on the right-hand side of (\ref{eq:500-2}) with a single Green function
without the profile function gives conventional  vacuum  spectrum  (\ref{eq:520-2}). It  should  be  noted
that in the situation with the initial particle impinging from infinity, the
last term in (\ref{eq:460-2}) gives zero contribution due to cancellation of
the sum of the contributions  from  the regions  $z_{1,2}<0$ and $z_{1,2}>0$
with  the contribution from the region $z_1<0$, $z_2>0$.

The  calculations  have  been  made  for  scalar  particles. The inclusion of
spin does not change the procedure of elimination of the 
indeterminacy $0\cdot \infty$. The results can be written in the same form 
in terms of the wavefunction for the pair $bc$.

\section*{Appendix B}

In this appendix, we consider formulas for the Green functions, which are
required for calculating the terms $I_{1,2}$ using  expressions
(\ref{eq:890-3}) and (\ref{eq:900-3}) 
in   the   oscillator approximation. In QCD, for quadratic parameterization of
the dipole cross section $\sigma_{q\bar{q}}(\rho)=C\rho^2$  (in terms of
quark transport coefficient, we have $C=\hat{q}/2n$), three-particle parton cross section $\sigma_{bc\bar{a}}$
 can also be written in quadratic form
\beq
\sigma_{bc\bar{a}}(\ro, \Rb)=
C_{b\bar{a}}(\ro_{b}-\ro_{\bar{a}})^2+C_{c\bar{a}}(\ro_{c}-\ro_{\bar{a}})^2
+C_{bc}(\ro_{b}-\ro_{c})^2\,.
\label{eq:1100}
\eeq
Here, $\ro=\ro_{b}-\ro_c$, $\Rb=x_c\ro_b+x_b\ro_c-\ro_{\bar{a}}$,
$\ro_{b}-\ro_{\bar{a}}=\Rb+x_c\ro$, and $\ro_{c}-\ro_{\bar{a}}=\Rb-x_b\ro$.
For process $q\to qg$ ($a=b=q$, $c=g$) we can obtain from (\ref{eq:670-2}) 
\beq
C_{bc}=C_{c\bar{a}}=\frac{9C}{8}\,,\,\,\,
C_{b\bar{a}}=-\frac{C}{8}\,.
\label{eq:1110}
\eeq
For the diagram in Fig. 2a, we have $\Rb=\ta_i=x_b\ta_f$. 
Introducing the new variable
\beq
\ub=\ro+\delb\,,\,\,\,
\delb=\ta_iB/C_3=x_b\ta_fB/C_3\,,
\label{eq:1120}
\eeq
where $B=x_cC_{b\bar{a}}-x_bC_{c\bar{a}}$,
$C_3=C_{b\bar{a}}x_c^2+C_{c\bar{a}}x_b^2+C_{bc}$,
we can write the expression for $\sigma_{bc\bar{a}}$ in the form 
\beq
\sigma_{bc\bar{a}}(\ro, \Rb)=
(A-B^2/C_3)\Rb^2+C_3\ub^2\,,
\label{eq:1130}
\eeq
where $A=C_{b\bar{a}}+C_{c\bar{a}}$.

With  allowance  for  relation  (\ref{eq:1130}),  the Hamiltonian
(\ref{eq:360-2}) for the system $bc\bar{a}$  as a function of $z$ 
can be written in terms of variable $\ub$ and vector $\ta_f$
in the form
\beq
H=H_{osc}-\frac{i d\theta(L-z)\ta_f^2}{2}+\frac{\epsilon^2}{2M}\,,
\label{eq:1140}
\eeq
where $d=nx_b^2(A-B^2/C_3)$, and $H_{osc}$ is  the  oscillator
Hamiltonian
\beq
H_{osc}=
-\frac{1}{2M}\,
\left(\frac{\partial}{\partial \ub}\right)^{2}
+\frac{M\Omega^2\ub^2}{2}
\label{eq:1150}
\eeq
with the complex frequency
\beq
\Omega=\sqrt{\frac{-inC_3\theta(L-z)}{M}}\,.
\label{eq:1160}
\eeq   
Note that $|\Omega|^2\propto \hat{q}$,
since  $C_3\propto C\propto \hat{q}$.
From (\ref{eq:1140}) one can see that  that the Green function ${\cal{K}}$ 
(for  the region  $z_1<L$ required  for  our  analysis)  
can be written in the form
\beq
{\cal{K}}(\ro_2,z_2|\ro_1,z_1)=K_{osc}(\ub_2,z_2|\ub_1,z_1)U(z_2,z_1)\,,
\label{eq:1170}
\eeq
\beq
U(z_2,z_1)=\exp{\left[-\frac{d\xi\ta_f^2}{2}
-\frac{i(z_2-z_1)\epsilon^2}{2M}\right]}\,,
\label{eq:1180}
\eeq
where $\xi=\mbox{min}(z_2,L)-z_1$, $\ub_i=\ro_i+\delb$,
and $K_{osc}$ is the Green  function  for  the oscillator  
Hamiltonian (\ref{eq:1150}), which can be written as
\beq
K_{osc}(\ub_2,z_2|\ub_1,z_1)=\frac{\gamma}{2\pi i}
\exp{\left[i(\alpha\ub_2^2+\beta \ub_1^2-\gamma\ub_1\cdot\ub_2)\right]}\,.
\label{eq:1190}
\eeq
Here, we have for $z_2<L$
\beq
\alpha=\beta=\frac{M\Omega}{2\tg{(\Omega(z_2-z_1))}}\,,\,\,\,\,\,
\gamma=\frac{M\Omega}{\sin{(\Omega(z_2-z_1))}}\,,
\label{eq:1200}
\eeq
and for configurations $z_2>L>z_1$
\beq
\alpha=\frac{M\Omega}{2[\tg{(\Omega\xi_1)}+\Omega\xi_2]}\,,\,\,
\beta=\frac{M\Omega[1-\Omega\xi_2\tg{(\Omega\xi_1}]}
{2[\tg{(\Omega\xi_1)}+\Omega\xi_2]}\,,\,\,\,\,\,
\gamma=\frac{M\Omega}
{\cos{\Omega\xi_1}[\tg{(\Omega\xi_1)}+\Omega\xi_2]}\,,
\label{eq:1210}
\eeq
where $\xi_1=L-z_1$, $\xi_2=z_2-L$.

In our formulas for spectra, differential operator  $\hat{g}$   is
acting  on  the  Green  function  ${\cal{K}}$  at  constant  value  of
$\ta_i$. Therefore, in $\hat{g}$  we  can  replace 
$\frac{\partial}{\partial \ro_2}\cdot\frac{\partial}{\partial \ro_1}$ by
$\frac{\partial}{\partial \ub_2}\cdot\frac{\partial}{\partial \ub_1}$.
Then, from (\ref{eq:1190}) one can readily obtain
\beq
\frac{\partial}{\partial \ro_2}\cdot   
\frac{\partial}{\partial \ro_1}
{\cal{K}}(\ro_2,z_2|\ro_1,z_1)=
-\left[2i\gamma+
(2\alpha\ub_2-\gamma\ub_1)\cdot(2\beta\ub_1-\gamma\ub_2)\right]
{\cal{K}}(\ro_2,z_2|\ro_1,z_1)\,.
\label{eq:1220}
\eeq
For the diagram in Fig. 2a, the Green function appears
for 
$\ro_1=0$, $\ro_2=\ta_f$, which corresponds to
\beq
\ub_{1,2}=\ta_f k_{1,2}\,,\,\,\, k_1=x_bB/C_3\,,\,\,\,
k_2=1+x_bB/C_3\,.
\label{eq:1230}
\eeq
Consequently,  for  $\ta_f=0$ that  appears  in  the  expressions for
$\langle p_\perp^2\rangle_{rad}$, we have $\ub_{1,2}=0$.
Then, considering the expressions (\ref{eq:660-2}) and (\ref{eq:1230}), 
we obtain
\beq
\hat{g}{\cal{K}}(\ro_2,z_2|\ro_1,z_1)\Big|_{\ro_{1,2}=\ta_f=0}=
\left(\frac{\alpha_sP_{ba}}{2M^2}\right)\cdot
\frac{\gamma^2}{\pi}\exp{\left[-\frac{i(z_2-z_1)\epsilon^2}{2M}\right]}\,.
\label{eq:1240}
\eeq
For  calculating  the term  $I_1$ (\ref{eq:890-3}), we must also know the
Laplacian in $\ta_f$ for $\ta_f=0$ of 
$\hat{g}{\cal{K}}(\ro_2,z_2|\ro_1,z_1)$ at $\ro_{2}=\ta_f$ and $\ro_1=0$.
The  right-hand  side  of  formula  (\ref{eq:1220}),
written as a function of $\ta_f$ for  $\ro_2=\ta_f$, $\ro_1=0$
has the form
\beq
-\frac{\gamma}{2\pi i}[2i\gamma +G\ta_f^2]
\exp\left[i\ta_f^2D-\frac{i(z_2-z_1)\epsilon^2}{2M}\right]\,,
\label{eq:1250}
\eeq
where
\beq
D=\alpha k_2^2+\beta k_1^2-\gamma k_1k_2+\frac{id\xi}{2}\,,
\label{eq:1260}
\eeq
\beq
G=(2\alpha k_2-\gamma k_1)(2\beta k_1-\gamma k_2)\,.
\label{eq:1270}
\eeq
Then, with  allowance  for  (\ref{eq:660-2}) and (\ref{eq:1250}),  
we  can easily obtain for $\ta_f=0$
\beq
\Delta^2
\hat{g}{\cal{K}}(\ro_2,z_2|\ro_1,z_1)\Big|_{\ro_2=\ta_f,\ro_1=0,\ta_f=0}=
\left(\frac{\alpha_sP_{ba}}{2M^2}\right)\cdot\frac{2\gamma
(2i\gamma D-G)}{i\pi}\exp{\left[-\frac{i(z_2-z_1)\epsilon^2}{2M}\right]}\,.
\label{eq:1280}
\eeq
For calculating analogs of formulas (\ref{eq:1240}) and (\ref{eq:1280}) for
the vacuum Green function, it is sufficient to set 
$d=0$ and replace the functions $\alpha$, $\beta$, and $\gamma$
 by their vacuum analogs
\beq
\alpha_0=\beta_0=\gamma_0/2=\frac{M}{2(z_2-z_1)}\,.
\label{eq:1290}
\eeq

For the virtual diagram in Fig. 2b in which the Green
function $\tilde{\cal{K}}$ appears, only the values of parameters 
$d$ and $k_{1,2}$ change in the resultant formula,
which are now defined as 
$d=n(A-B^2/C_3)$ and $k_{1,2}=B/C_3$.


\begin{thebibliography}{99}

\bibitem{Heinz_hydro}
U.W. Heinz,
Landolt-Bornstein {\bf 23}, 240 (2010)
[arXiv:0901.4355].


\bibitem{Heinz_tau}
H. Song, S.A. Bass, U. Heinz, and T. Hirano,
Phys. Rev. C{\bf 83}, 054910 (2011) , Erratum: Phys.
Rev. C{\bf 86}, 059903 (2012)
[arXiv:1101.4638].



\bibitem{Pasechnik}
R. Pasechnik and M.  Šumbera,
Universe {\bf 3}, 7 (2017) 
[arXiv:1611.01533].



\bibitem{Wied_JQ}
U.A. Wiedemann,
Landolt-Bornstein {\bf 23}, 521 (2010)
[arXiv:0908.2306].

\bibitem{Bjorken1} J.D.~Bjorken, Fermilab preprint 
82/59-THY (1982, unpublished).



\bibitem{GW}
M.~Gyulassy and X.N.~Wang,
Nucl. Phys. B{\bf 420}, 583 (1994)
[nucl-th/9306003].


\bibitem{BDMPS1}
R.~Baier, Y.L.~Dokshitzer, A.H.~Mueller, S.~Peign\'e, and D.~Schiff,
Nucl.\ Phys.\ B{\bf 483}, 291 (1997) [hep-ph/9607355].


\bibitem{BDMPS2}
R.~Baier, Y.L.~Dokshitzer, A.H.~Mueller, S.~Peign\'e, and D.~Schiff,
Nucl.\ Phys.\
B{\bf 484}, 265 (1997) [hep-ph/9608322].

\bibitem{LCPI1}
B.G.~Zakharov, JETP\ Lett. {\bf 63}, 952 (1996)
[hep-ph/9607440]. 




\bibitem{GLV1}
M.~Gyulassy, P.~L\'evai, and I.~Vitev, 
Nucl.\ Phys. B{\bf 594}, 371 (2001) [hep-ph/0006010].

\bibitem{AMY}
P.~Arnold, G.D.~Moore, and L.G.~Yaffe,
JHEP {\bf 0206}, 030 (2002) [hep-ph/0204343].


\bibitem{W1}
U.A.~Wiedemann,
Nucl.\ Phys.\ A{\bf 690}, 731 (2001 [hep-ph/0008241].

\bibitem{Z_coll}
B.G.~Zakharov,
JETP Lett. {\bf 86}, 444 (2007)
[arXiv:0708.0816].


\bibitem{Gale_coll}
G.-Y. Qin, J. Ruppert, C. Gale, S. Jeon, G.D. Moore, and M.G. Mustafa,
Phys.~Rev.~Lett. {\bf 100}, 072301 (2008) 
[arXiv:0710.0605].

\bibitem{LP}
L.D. Landau and I.Ya. Pomeranchuk,
{Dokl. Akad. Nauk SSSR} {\bf 92}, 535, 735 (1953).

\bibitem{Migdal}
A.B. Migdal, {Phys. Rev.} {\bf 103}, 1811 (1956).



\bibitem{Z_OA}
B.G. Zakharov,
JETP Lett. {\bf 73}, 49 (2001)
[hep-ph/0012360].

\bibitem{Arnold_2g1}
P. Arnold and S. Iqbal,
JHEP {\bf 1504}, 070 (2015), Erratum: JHEP {\bf 1609}, 072 (2016)
[arXiv:1501.04964].

\bibitem{BDMS_RAA}
R.~Baier, Yu.L.~Dokshitzer, A.H.~Mueller, and
D.~Schiff, JHEP {\bf 0109}, 033 (2001) [hep-ph/0106347].

\bibitem{RAA13} 	
B.G.~Zakharov, 
J. Phys. G{\bf 40}, 085003  (2013) [arXiv:1304.5742].


\bibitem{RPP14}
B.G.~Zakharov, 
J. Phys. G{\bf 41}, 075008  (2014) 
[arXiv:1311.1159].

\bibitem{Mueller_dijet}
A.H. Mueller, B. Wu, B.-W. Xiao, and F. Yuan,
Phys.~Lett. {\bf B}763, 208 (2016)
[arXiv:1604.04250].

\bibitem{STAR1}
L. Adamczyk {\it et al.}  [STAR Collaboration],
Phys.Rev. C{\bf 96}, 024905 (2017)
[arXiv:1702.01108].


\bibitem{ALICE_hjet}
J. Norman [for ALICE Collaboration],
arXiv:1901.02706.

\bibitem{Gyulassy_dijet}
M. Gyulassy, P. Levai, J. Liao, S. Shi, F. Yuan, and X.N. Wang,
Nucl. Phys. {\bf A}982, 627 (2019)
[arXiv:1808.03238].


\bibitem{Wu}
B. Wu, JHEP {\bf 1110}, 029 (2011)
[arXiv:1102.0388].

\bibitem{Mueller_pt}
T. Liou, A.H. Mueller, and B. Wu,
Nucl.~Phys. A{\bf 916}, 102 (2013) 
[arXiv:1304.7677].

\bibitem{Blaizot_pt} 	
J.-P. Blaizot and Y. Mehtar-Tani,
Nucl.~Phys. A{\bf 929}, 202 (2014)
[arXiv:1403.2323].




\bibitem{LCPI_PT}
B.G.~Zakharov, JETP\ Lett. 
{\bf 70}, 176 (1999) [hep-ph/9906536].


\bibitem{BSZ}
R. Baier, D. Schiff, and B.G. Zakharov,
Ann.~Rev.~Nucl.~Part.~Sci. {\bf 50}, 37 (2000)
[hep-ph/0002198].

\bibitem{Z_NP05}
B.G. Zakharov,
Nucl.~Phys.~Proc.~Suppl. {\bf 146}, 151 (2005)
[hep-ph/0412117].


\bibitem{LL4} V.B. Berestetski, E.M. Lifshits, and L.P. Pitaevski,
{\it Quantum Electrodynamics (Landau Course of 
Theoretical Physics Vol. 4)}, Oxford, Pergamon
  Press, 1979. 


\bibitem{Z_kinb}
B.G. Zakharov,
JETP Lett. {\bf 80}, 76 (2004)
[hep-ph/0406063].


\bibitem{FH}
R.P. Feynman and A.R. Hibbs,
{\sl Quantum Mechanics and
Path  Integrals}, McGRAW–HILL Book Company, New
York 1965



\bibitem{Zpath87}
B.G.~Zakharov,
Sov. J. Nucl. Phys. {\bf 46}, 92 (1987).

\bibitem{BKS}
J.D. Bjorken, J.B. Kogut, and D.E. Soper,
Phys. Rev. D{\bf 3}, 1382 (1971).

\bibitem{B-L}
G.P. Lepage and S.J. Brodsky,
Phys. Rev. D{\bf 22}, 2157 (1980).





\bibitem{NZ_SIGMA3}
N.N. Nikolaev and B.G. Zakharov,
Z. Phys. C{\bf 64}, 631 (1994)
[hep-ph/9306230].

\bibitem{SLAC1}
B.G. Zakharov,
JETP Lett. {\bf 64}, 781 (1996)
[hep-ph/9612431].


\bibitem{AZ}
P. Aurenche and B.G. Zakharov,
JETP Lett. {\bf 85}, 149 (2007) 
[hep-ph/0612343].


\bibitem{Baier_q}
R. Baier,
Nucl. Phys. A{\bf 715}, 209 (2003)
[hep-ph/0209038].

\bibitem{JET_q}
K.M. Burke {\it et al.} [JET Collaboration]
Phys.~Rev. C{\bf 90}, 014909  (2014) 
[arXiv:1312.5003].


\bibitem{LH}
P.~L\'evai and U.~Heinz,
Phys.\ Rev.\ C{\bf 57}, 1879 (1998) [hep-ph/9710463].

\bibitem{Bielefeld_Md}
O.~Kaczmarek and F.~Zantow,
Phys. Rev. D{\bf 71}, 114510 (2005) [hep-lat/0503017].



\bibitem{Bjorken}
J.D.~Bjorken, 
Phys.\ Rev. D{\bf 27}, 140 (1983).



\bibitem{LCPI_YF98}
B.G. Zakharov,
Phys. Atom. Nucl. {\bf 61}, 838 (1998)
[hep-ph/9807540].

\bibitem{NZ_HERA}
N.N.~Nikolaev and B.G.~Zakharov,
Phys. Lett. B{\bf 327}, 149 (1994) [hep-ph/9402209]. 

\bibitem{shuryak1}
E.V.~Shuryak, Rev.\ Mod.\ Phys. {\bf 65}, 1 (1993).

\bibitem{Z_TM}
B.G. Zakharov, JETP Lett. {\bf 76}, 201 (2002)
[hep-ph/0207206].

\bibitem{DKT}
Yu.L.~Dokshitzer, V.A.~Khoze, and S.I.~Troyan,
Phys.\ Rev. D{\bf 53}, 89 (1996) [hep-ph/9506425]. 



\end{thebibliography}
\end{document}